\documentclass[%
 reprint,
 amsmath,amssymb,
 aps,
pra,
]{revtex4-1}
\usepackage{graphicx}
\usepackage{dcolumn}
\usepackage{bm}
\usepackage{textcomp}
\usepackage{bigstrut,multirow,rotating,amsmath}

\usepackage[dvipdfm,colorlinks,linkcolor=blue, urlcolor=blue, anchorcolor=blue, citecolor=blue]{hyperref}
\begin{document}


\title{Analytically solvable many-body Rosen-Zener quantum battery}

\author{Wei-Xi Guo}
\affiliation{College of Physics and Electronic Engineering, Northwest Normal University, Lanzhou, 730070, China}
\author{Fang-Mei Yang}
\affiliation{College of Physics and Electronic Engineering, Northwest Normal University, Lanzhou, 730070, China}
\author{Fu-Quan Dou}
\email{doufq@nwnu.edu.cn}
\affiliation{College of Physics and Electronic Engineering, Northwest Normal University, Lanzhou, 730070, China}

\begin{abstract}
Quantum batteries are energy storage devices that satisfy quantum mechanical principles. How to obtain analytical solutions for quantum battery systems and achieve a full charging is a crucial element of the quantum battery. Here, we investigate the Rosen-Zener quantum battery with $N$ two-level systems, which includes atomic interactions and external driving field. The analytical solutions of the stored energy, changing power, energy quantum fluctuations, and von Neumann entropy (diagonal entropy) are derived by employing the gauge transformation. We demonstrate that full charging process can be achieved when the external driving field strength and scanning period conforms to a quantitative relationship. The local maximum value of the final stored energy corresponds to the local minimum values of the final energy fluctuations and diagonal entropy. Moreover, we find that the atomic interaction induces the quantum phase transition and the maximum stored energy of the quantum battery reaches the maximum value near the quantum phase transition point. Our result provides an insightful theoretical scheme to realize the efficient quantum battery.
\end{abstract}

\maketitle


\section{Introduction}\label{section1}
With the decline of fossil fuels and the aggravation of global energy crisis, there is a constant search for alternative energy sources \cite{PhysRevResearch.2.023113}. In this context, the growth of renewable energies makes the issue of energy storage extremely urgent. Likewise, with the boost of quantum thermodynamics \cite{PhysRevA.99.062111, PhysRevLett.125.180603,Mukherjee_2021,PhysRevLett.124.210603,PhysRevResearch.2.043247,PhysRevE.90.012119} and the growing demand for device miniaturization \cite{Millen_2016}, the size of energy storage devices has approached molecular or even atomic scale. It is necessary to consider the role of quantum effects on energy storage \cite{PhysRevResearch.2.023113,
PhysRevA.100.043833, centrone2021charging, PhysRevResearch.4.013172}. Scientists have tried to exploit quantum system to create a new class of batteries with ultra-higher energy density, ultra-compact size, ultra-fast charging and ultra-long life time \cite{centrone2021charging}. With these requirements, Alicki and Fannes first proposed the concept of quantum battery (QB), i.e., a quantum system that stores or supplies energy \cite{PhysRevE.87.042123, campaioli2023colloquium}.
Previous researches have shown the importance of quantum features in improving the performances of QB, ranging from energy storage \cite{PhysRevA.103.033715, Dou_2020, PhysRevResearch.2.023095, PhysRevApplied.14.024092, PhysRevB.98.205423, PhysRevE.99.052106, Crescente_2020}, work extraction \cite{Dou2022charging, e23050612, Mitchison2021chargingquantum, PhysRevLett.122.047702, PhysRevLett.125.040601, PhysRevE.102.052109, e23121627, Barra_2022, PhysRevB.104.245418, PhysRevE.104.064143, Bhattacharjee2021, PhysRevA.107.032218, PhysRevA.107.032218}, charging power \cite{PhysRevA.106.032212,PhysRevResearch.2.023095, Bhattacharjee2021, Binder_2015, PhysRevA.97.022106, PhysRevB.102.245407, PhysRevB.105.115405, yang2023threelevel, PhysRevLett.118.150601, PhysRevLett.120.117702, Carrega_2020, PhysRevLett.125.236402, Perarnau_Llobet_2019, PhysRevA.103.052220, PhysRevApplied.14.024092, PhysRevE.104.054117, PhysRevA.106.022618} to energy quantum fluctuations \cite{Friis2018precisionwork, PhysRevLett.125.040601, PhysRevE.98.032132,PhysRevA.107.022215}.

A key challenge is how to obtain analytical solutions of the QB system and ensure the stability of the charging or energy transfer process. There are many efforts to derive analytical solutions of single- \cite{Crescente_2020, e23050612, PhysRevE.102.042111, PhysRevE.102.062133, Crescente_2023} or many-body QB \cite{PhysRevE.99.052106, chen2020, Zhang2023, Caravelli2021energystorage, Carrega_2020} by using various 
 approximation methods. In addition, stable charging requires a control protocol to bring the battery's charge to a stationary value and removes the need for precisely timed switching of the battery-charger coupling \cite{PhysRevE.100.032107, PhysRevE.101.062114, Mitchison2021chargingquantum}. To this end, various schemes have been devised, including the quantum feedback control method \cite{Mitchison2021chargingquantum}, the transitionless quantum driving \cite{PhysRevE.100.032107, PhysRevE.101.062114}, the shortcut to adiabaticity \cite{Dou2021}, and the optimal control \cite{PhysRevA.107.032218}.

The two-level systems (TLSs) are a favorite model in many areas of physics, and are successful in describing a large variety of physical phenomena \cite{Dou_2016, PhysRevA.89.012123, PhysRevA.93.043419, PhysRevA.98.022102}.
The original theoretical proposal of QB is based on TLSs \cite{Binder_2015}. Later on, there were a number of works discussed on two-level QB \cite{Dou2022charging, PhysRevLett.111.240401, PhysRevE.102.052109, PhysRevLett.118.150601, PhysRevLett.125.040601, PhysRevE.99.052106, chen2020, Zhang2023, Crescente_2020, PhysRevB.98.205423, PhysRevB.99.205437, Mitchison2021chargingquantum, moraes2022charging, PhysRevE.100.032107, PhysRevLett.127.100601, PhysRevB.100.115142, Binder_2015,PhysRevE.94.052122, PhysRevLett.120.117702, PhysRevB.105.115405, PhysRevB.102.245407, PhysRevA.97.022106, PhysRevLett.122.047702, e23050612, Superabsorption2022, PhysRevA.104.042209, PhysRevA.102.060201, Caravelli2021energystorage, Carrega_2020, PhysRevE.102.042111, PhysRevE.102.062133}.
Rosen-Zener (RZ) model is a typical model among the TLSs which is first proposed to study the spin-flip of two-level atoms interacting with a rotating magnetic field to account for the double Stern-Gerlach experiments \cite{PhysRev.40.502}. This model has extensive applications in quantum coherence control \cite{PhysRevA.76.053404}, ultra-cold atoms molecule transformation \cite{PhysRevA.78.043609,ISHKHANYAN2009218}, quantum interference \cite{PhysRevA.80.013619, PhysRevA.78.063621,PhysRevA.101.023618, PhysRevA.102.033323}, superconductivity \cite{PhysRevLett.99.205303}, quantum information \cite{PhysRevLett.105.090502}, and quantum tunneling \cite{PhysRevA.77.013402}, where the energy bias between two levels is fixed and the coupling is time dependent. Recently, QB has been proposed using a time-dependent driving field as a charger \cite{Crescente_2020, PhysRevE.99.052106, chen2020}, such as harmonic \cite{PhysRevE.99.052106, Crescente_2020}, general harmonic \cite{chen2020}, and rectangular pulses \cite{Crescente_2020}. Besides, the simulation of a time-dependent driven two-level QB on an IBM quantum chips has been realized \cite{batteries8050043}.
However, most of the associated theoretical discussions about two-level QB are limited to noninteracting atoms. In fact, the actual physics system always involves interatomic interactions. In addition, the collective behavior from $N$ two-level systems has been studied for quantum heat engine \cite{PhysRevLett.124.210603}. It is a quite natural question to study the effects of both many-body and the atomic interaction on the charging performance \cite{PhysRevE.99.052106}.
Although the external driving as a charger has been studied numerically and analytically \cite{Crescente_2020, PhysRevE.99.052106, chen2020}, it is still difficult to give the analytical expression and to achieve a stable charging of the many-body QB considering the interaction between atoms.

In this paper, we investigate the many-body RZ quantum battery with both atomic interactions and external driving field. The analytic results of the stored energy, charging power, energy quantum fluctuations, and von Neumann entropy (diagonal entropy) are derived using the gauge transformation \cite{WANG1993189, WANG199413, PhysRevA.98.022136}. These results are then compared with numerical calculations.
In addition, we also determine the conditions to achieve full charging and obtain the qualitative relationship among the final stored energy, the final energy fluctuations, and the final diagaonal entropy. The effects of the atomic interaction on the charging performance of the QB have been further considered. Finally, we also simulate the dependence of the QB's stored energy, charging power, energy fluctuations and diagonal entropy on the number of TLSs.

The remaining of this paper is organized as follows. In Sec. \ref{section2}, we show the charging protocols of the RZ quantum battery, while in Sec. \ref{section3}, the analytical solutions of the QB are derived. The relationship among the stored energy, energy quantum fluctuations, and diagonal entropy is obtained in Sec. \ref{section4}. In Sec. \ref{section5}, we analyze the role of the
atomic interactions and the number of atoms. Finally, the conclusions are given in Sec. \ref{section6}.

\section{THE QUANTUM BATTERY model}\label{section2}
\begin{figure}[tbp]
 \centering
 \includegraphics[width=0.485\textwidth]{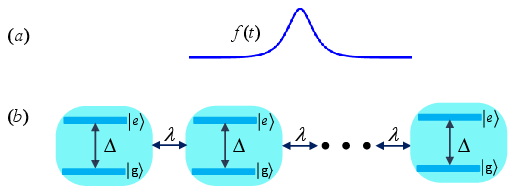}
 \caption{A sketch of the RZ quantum battery. (a) An external driving field $f(t)$. (b) A set of $N$ identical TLSs with atomic interactions. During the charging time $0<t<\tau$ $(\tau= T)$, the QB is coupled with the driving field and atomic interactions.}
\label{fig1}
\end{figure}

We consider a QB model as an ensemble of $N$ TLSs, which are charged by an external driving field and atomic interaction, as sketched in Fig. \ref{fig1}. The Hamiltonian of the QB system is
\begin{equation}
H(t)=H_0+\Theta(t)\left[H_1(t)+H_{a-a}\right],
\label{H}
\end{equation}
where the time-dependent parameter $\Theta(t)$ describes the charging time interval, which is given by a step function equal to $1$ for $t\in[0,\tau]$ and zero elsewhere. $H_0$ describes the time independent TLSs of the QB. $H_{1}$ represents the external driving field and $H_{a-a}$ is the interactions between atoms, and with the following forms:
\begin{equation}
H_0=\frac{\Delta}{2}\sum_{i=1}^{N}\hat{\sigma}_i^z=\Delta\hat{J}_z,
\end{equation}
\begin{equation}
H_1(t)=\frac{f(t)}{2}\sum_{i=1}^{N}\hat{\sigma}_i^x=f(t)\hat{J}_x,
\end{equation}
\begin{equation}
\begin{aligned}
H_{a-a}=\frac{\eta}{2N}\sum_{i\neq j}^{N}\hat{\sigma}_i^z\hat{\sigma}_j^z=\frac{2\eta}{N}\hat{J}_z^2.
\end{aligned}
\end{equation}
Here $\Delta=\hbar\omega_0$ denotes the energy level gap between the ground state $|g\rangle$ and the excited state $|e\rangle$, respectively. The coupling $f(t)$ is time-dependent external driving field, which can take various forms, such as Gaussian \cite{PhysRevA.27.2744}, exponential \cite{PhysRevA.49.265}, and hyperbolic secant \cite{PhysRevA.89.043411}. In this work, we choose \cite{PhysRevA.77.013402}
\begin{equation}
f(t) = \begin{cases}
  0, & t < 0, t > T, \\
  v_0 \sin^2 \left( \dfrac{\pi t}{T} \right), & t \in [0, T],
\end{cases}
\end{equation}
where $v_{0}, T$ represent the strength and the scanning period of the external driving field, respectively. $\hat{\sigma}_\alpha$ $(\alpha=x,y,z)$ are the usual Pauli matrices. $\eta$ is the atom-atom interactions strength including the repulsive $(\eta > 0)$ and attractive $(\eta < 0)$ interactions, and we define the scaled interactions strength $\lambda=\eta/\Delta$. The collective atom operators $\hat{J}_\alpha=\sum_{i=1}^{N}\frac{1}{2}\hat{\sigma}_i^\alpha$. The Hamiltonian of the QB system can be described by the Dicke states $\left |s,m \right \rangle$ $(m=-s,-s+1,\cdots,s)$. In all calculations, we take $\omega_0$ as the dimensionless parameter and set $\omega_0=1$.

It should be noted that if a two-level atom is viewed as a $1/2$-spin, the model can be regarded as the driven transverse-field Ising model, or as the Ising model with longitudinal and transverse fields \cite{dutta_aeppli_chakrabarti_divakaran_rosenbaum_sen_2015,PhysRevLett.124.210603,PhysRevResearch.2.043247}. As uncharged state, the QB is prepared in the ground state of $N$ TLSs. Thus, the initial state of the QB system is
$\left | \psi(0) \right\rangle=\left |N/2,-N/2\right\rangle$. The wave function evaluates according to the
Schr\"{o}dinger equation $i\hbar\partial \left| {\psi (t)} \right\rangle / \partial t = H (t)\left| {\psi (t)} \right\rangle$.

During charging, the stored energy of QB is
\begin{equation}
\label{E}
E(t)=\left \langle \psi(t) \right |H_0\left | \psi(t) \right \rangle-\left \langle \psi(0)
\right | H_0 \left | \psi(0) \right \rangle,
\end{equation}
the average charging power is
\begin{equation}
\label{P}
P(t)=\frac{E(t)}{t},
\end{equation}
and the instantaneous charging power is
\begin{equation}
\label{PI}
P_{I}(t)=\mathrm{tr}\left[ {{H_0}\frac{{d\rho (t)}}{{dt}}} \right],
\end{equation}
where the density matrix $\rho(t)=\left | \psi(t) \right \rangle \left \langle \psi(t) \right |$.

The knowledge of the stored energy and charging power as a function of time is not sufficient to fully characterize the performance of the QB \cite{Crescente_2020}. Indeed, together with this, it is important to have information about the energy quantum fluctuations \cite{Friis2018precisionwork, Crescente_2020,PhysRevLett.125.040601} and von Neumann entropy \cite{PhysRevB.104.245418}. Therefore, we also define the energy quantum fluctuations and von Neumann entropy (the diagonal entropy) \cite{PhysRevLett.113.140401} as follows:
\begin{equation}
\label{sig}
\Sigma(t)=\sqrt{\langle H_0^2(t) \rangle-(\langle H_0(t) \rangle)^2},
\end{equation}
\begin{equation}
\label{s}
S(t)=- \mathrm{tr}\left[ {\rho_{diag} (t)\log_2 \rho_{diag}(t)}\right]=-\sum_{i}{\rho_{ii} (t)\log_2 \rho_{ii}(t)},
\end{equation}
where $\rho_{diag}$ denotes the state obtained from $\rho(t)$ by taking diagonal elements (i.e., deleting all off-diagonal elements) \cite{PhysRevLett.113.140401} and $\rho(t)$ is the density matrix of the whole system. The von Neumann entropy (\ref{s}) as defined above is also called the diagonal entropy \cite{PhysRevLett.129.130602}. Notice that no entanglement exists in the system. So the diagonal entropy is precisely the coherence. (The relative entropy of coherence is defined as $C=S(t)-S_{vN}(t)$, here $S_{vN}=-\mathrm{tr}\left[ {\rho(t)\log_2 \rho(t)}\right]$ \cite{PhysRevLett.113.140401}. Without loss of generality, we usually choose the maximum stored energy $E_{max}$, maximum average charging power $P_{max}$, maximum energy fluctuations $\Sigma_{max}$, maximum diagonal entropy $S_{max}$, final stored energy $E(\tau)$, final energy fluctuations $\Sigma(\tau)$, and final diagonal entropy $S(\tau)$ to measure QB's performances.
\section{Analytical solutions of the QB}\label{section3}
In this section, we use the gauge transformation to obtain the analytical results and analyze the charging performance of the QB. The Lie algebraic structure of the Hamiltonian in our time-dependent driven quantum systems suggests that gauge transformation can be a potent method to solve the Schr\"{o}dinger equation \cite{WANG1993189, WANG199413}. The gauge transformation method has been widely used in various models, including the Landau-Zener model \cite{LI20181}, Allen-Eberly model \cite{Li2018}, and Russell-Saunders coupled model \cite{PhysRevA.58.3328}. The analytical solutions of the many-body RZ quantum battery with the atomic interaction can be obtained by the following steps:

First, we apply two unitary transformations to the many-body RZ model in succession, where the unitary matrices are

\begin{equation}
\begin{aligned}
U_1(t)&=e^{i \mu (t)\hat{J}_x},\\
U_2(t)&=e^{i \nu (t)\hat{J}_y}.
\end{aligned}
\end{equation}
Here $\mu(t)$ and $\nu(t)$ are time-dependent undetermined functions.

Then, the transformed Hamiltonian becomes as follows:
\begin{equation}
\begin{aligned}
H'(t)&= U_1(t)H(t)U_1^\dag(t)-iU_1(t)\frac{d}{dt}U_1^\dag(t)\\
&=\left(v_0\sin^2(\frac{\pi t}{T})+\dot{\mu}(t)\right)\hat{J}_x +\sin\mu(t)\hat{J}_y+\cos\mu(t)\hat{J}_z\\
&+\frac{2\lambda }{N}\left[ \frac{1}{2}\sin2\mu(t)\left({\hat{J}_z}{\hat{J}_y}+{\hat{J}_y}{\hat{J}_z}\right)
+ {\sin^2\mu(t) \hat{J}_y^2}\right]\\
&+\frac{2\lambda }{N}\left[\cos^2\mu(t)\hat{J}_z^2\right],
\end{aligned}
\end{equation}
\begin{equation}
\begin{aligned}
H''(t)&= U_2(t)H'(t)U_2^\dag(t)-iU_2(t)\frac{d}{dt}U_2^\dag(t)\\
&=A_1(t)\hat{J}_x+A_2(t)\hat{J}_y+\cdots+A_{12}(t)\hat{J}_z^2,
\end{aligned}
\end{equation}
the above Hamiltonian $H''(t)$ is comprised of twelve terms, where the operators are represented by $\hat{J}_\alpha$ and $\hat{J}_\alpha\hat{J}_\beta$ with $\alpha,\beta=x,y,z$. The coefficients associated with each operator are denoted as $A_n(t)$ where $n=1,2,\dots,12$ (see Appendix \ref{app1}). The aim of the transformation described above is to obtain a suitable set of values for $\mu$ and $\nu$. By setting $\nu(t)=\pi$ and $A_1=0$, we can derive the analytical expressions for $\mu(t)$ with the initial condition $\mu(0)=\pi$,
\begin{equation}
\mu (t)=\pi - {\frac{{{v_0}t}}{2} + \frac{{{v_0}T}}{{4\pi }}\sin \left( {\frac{{2\pi t}}{T}} \right)}.
\end{equation}

Thus, the evolution operator of the system is
\begin{widetext}
\begin{equation}
\begin{aligned}
\label{U3}
U_3(t)=\mathcal{T}e^{-i\int_0^t {H''(t)dt}}\approx e^{-i\int_0^t \left( A_2(t)\hat J_y+A_3(t)\hat J_z + A_8(t)\hat J_y\hat J_z+A_9(t)\hat J_z\hat J_y + A_{11}(t)\hat J_y^2+A_{12}(t)\hat J_z^2 \right)dt},
\end{aligned}
\end{equation}
\end{widetext}
here $\mathcal{T}$ is a time-ordering operator. The symbol `$\approx$' is used because $H''$ is time-dependent (the equals sign is only valid if the Hamiltonian $H''$ commutes with different times). To obtain a high charging power, the time in our calculation is taken very short. The Hamiltonian basically satisfies the commutation relation, i.e., $H''(t)$ commutes with $H''(t')$. The agreement between the analytical results and the exact numerical calculations 
 will further show that this approximation is reasonable.

The direct integration of the composite trigonometric function in Eq. (\ref{U3}) is complicated, and thus we replace it with a Bessel function for the integration operation. To obtain the high average charging power of the QB, we consider a short scanning period $T$, i.e., a high driving field frequency. This results in a negligible contribution to the integral from the $n\geq2$ part of the Bessel functions $J_n$. Therefore, the time-dependent state of $H(t)$-system can be expressed as
\begin{widetext}
\begin{equation}
\begin{aligned}
\left |\psi(t) \right \rangle =U_1^\dagger(t) U_2^\dagger(t) U_3(t) \left | \psi(0) \right \rangle=e^{-i \mu (t)\hat{J}_x} e^{-i \nu (t)\hat{J}_y} e^{i \left[B_2\hat{J}_y + B_3\hat J_z+B_8\hat J_y\hat J_z+B_9\hat J_z\hat J_y + B_{11}\hat J_y^2+B_{12}\hat J_z^2\right]}\left | \frac{N}{2},-\frac{N}{2} \right \rangle,
\end{aligned}
\end{equation}
\end{widetext}
where the expressions of $B_{n}$ $(n=1,2,\cdots,12)$ are shown in Appendix \ref{app2}.

After substituting the wave function $\left|\psi(t)\right\rangle$ into Eqs. (\ref{E}), (\ref{P}) and (\ref{PI}), the stored energy, average charging power and instantaneous charging power of the QB are given by
\begin{equation}
\begin{aligned}
\label{Et}
E(t)=& \frac{{N{\Delta}}}{2}\left[ {1 + \cos \mu \left( t \right)} \right] +  {\frac{{{N^2}\left( {{B_8} + {B_9}} \right)}}{4}}\\
\times& \frac{{{B_2}\cos \mu \left( t \right) + {B_3}\sin \mu \left( t \right)}}{{B_2^2 + B_3^2}}\left( {\cos \sqrt {B_2^2 + B_3^2}  - 1} \right),
\end{aligned}
\end{equation}
\begin{equation}
\begin{aligned}
\label{Pt}
P(t)=& \frac{{N{\Delta}}}{2t}\left[ {1 + \cos \mu \left( t \right)} \right] +  {\frac{{{N^2}\left( {{B_8} + {B_9}} \right)}}{4t}}\\
\times& \frac{{{B_2}\cos \mu \left( t \right) + {B_3}\sin \mu \left( t \right)}}{{B_2^2 + B_3^2}}\left( {\cos \sqrt {B_2^2 + B_3^2}- 1} \right),
\end{aligned}
\end{equation}

\begin{equation}
\begin{aligned}
\label{PIt}
{P_I}(t) &= \frac{{N{\Delta}{v_0}}}{2}{\sin ^2}\left( {\frac{{\pi t}}{T}} \right)\sin \mu (t)\\
 &- {v_0}{\sin ^2}\left( {\frac{{\pi t}}{T}} \right)\frac{{\left[ {{B_3}\cos \mu (t) - {B_2}\sin \mu (t)} \right]}}{{4\left( {B_2^2 + B_3^2} \right)}}\\
 &\times {N^2}\left( {{B_8} + {B_9}} \right)\left( {\cos \sqrt {B_2^2 + B_3^2} - 1} \right).
\end{aligned}
\end{equation}

The energy quantum fluctuations of the battery can be written as
\begin{equation}
\label{sig}
\Sigma(t) = \sqrt {\frac{N\Delta}{4} - \frac{ \left[ E(t)-\dfrac{N \Delta}{2} \right]^2} {N}},
\end{equation}
the detailed calculation is shown in Appendix \ref{app3}.

The expression of the diagonal entropy is obtained by substituting $\rho(t)$ into Eq. (\ref{s}),
\begin{equation}
\begin{aligned}
\label{S}
&S(t)=\left| {\sin \mu (t)} \right|\times\\
&{\log _2}\left[ {\sqrt {(N - 1){e^{\frac{\pi}{2}}}} \left( {1 - {B_8}{B_9}\left( {{B_2}\cos \mu (t) - {B_3}\sin \mu (t)} \right)} \right)} \right].
\end{aligned}
\end{equation}
\begin{figure}[htbp]
 \centering
 \includegraphics[width=0.485\textwidth]{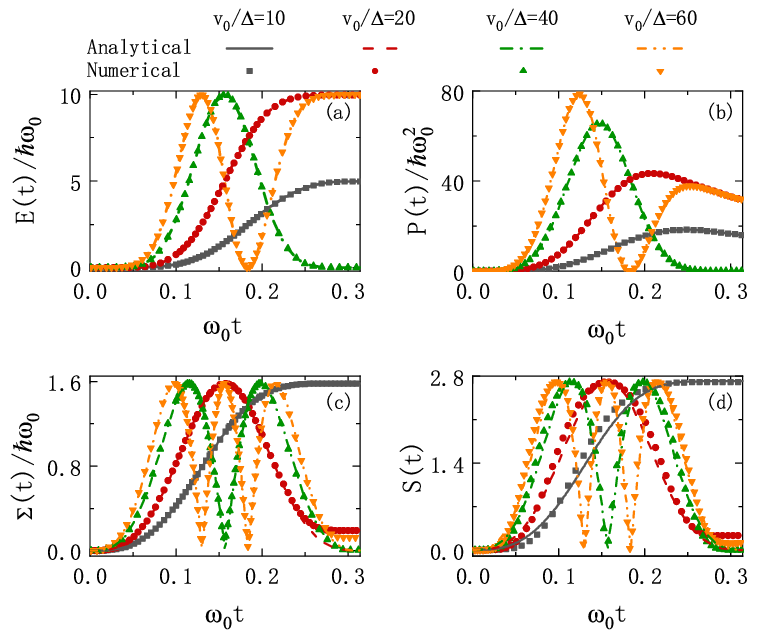}
\caption{Behaviors of (a) stored energy $E(t)$ (in units of $\hbar\omega_0$), (b) average charging power $P(t)$ (in units of $\hbar\omega_0^2$), (c) quantum fluctuations $\Sigma(t)$ (in units of $\hbar\omega_0$) and (d) diagonal entropy $S(t)$ as a function of $\omega_0 t$ for different driving field strength. Symbols represent numerical results, while lines depict analytical results. The external driving field strength are $v_0/\Delta=10$ (black solid lines and squares), $v_0/\Delta=20$ (red dashed lines and circles), $v_0/\Delta=40$ (green dash-dotted lines and up-triangles), and $v_0/\Delta=60$ (orange double dash-dotted lines and down-triangles). Other parameters: $\lambda=2, N=10$, and $\omega_0T=0.1\pi$.}
\label{fig2}
\end{figure}

We compare the analytical and numerical results for the stored energy, average charging power, quantum fluctuations, and diagonal entropy as functions of $\omega_0t$. The results of the comparison are shown in Fig. \ref{fig2} for different driving field strength. All numerical results are obtained by a numerically exact solution of the Schr\"{o}dinger equation with Hamiltonian (\ref{H}), without approximations. It is evident that the analytical results are in good agreement with the numerical results. Within the interval $t\in[0,T]$, higher driving field strength results in better charging performance of the QB, i.e., faster charging. This means the peak in stored energy is earlier for higher driving field strength.
The instantaneous charging power is also displayed in Eq. ({\ref{PIt}}) and represents the slope of stored energy.
The sign of the instantaneous charging power provides information about the energy transfer direction. Specifically, positive values represent energy transfer from the charger to the battery, whereas negative values suggest energy flow back from the battery to the charger. The quantum fluctuations and diagonal entropy represent the uncertainty of stored energy and the coherence, respectively. Remarkably, the evolution of the uncertainty and the coherence is almost synchronized.
An enhancement in coherence accompanies an increase in energy for lower driving field strength.
In contrast, under higher drive field strength, the extent of coherence increases while the speed of energy transfer also accelerates.
It indicates that the coherence, as an important resources, promotes the energy transfer between them.

\begin{figure}[tbp]
 \centering
 \includegraphics[width=0.485\textwidth]{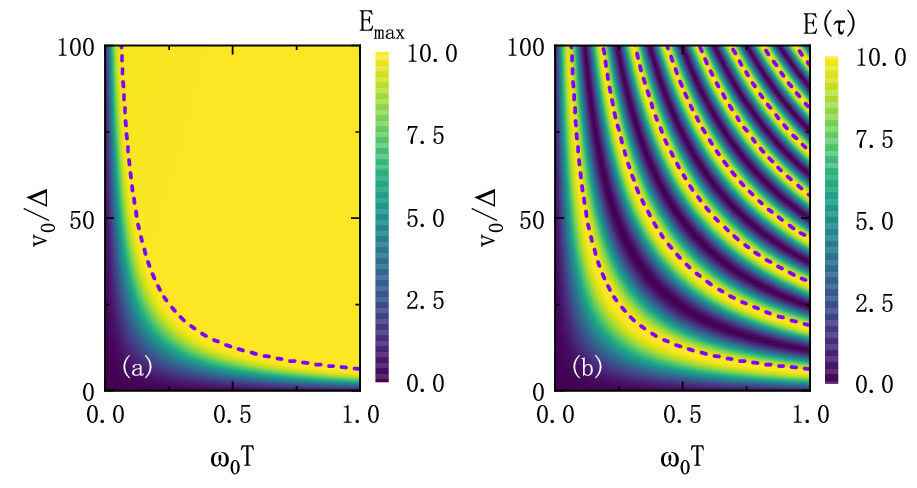}
\caption{Contour plots of QB's (a) maximum stored energy $E_{max}$ (in units of $\hbar\omega_0$) and (b) final stored energy $E(\tau)$ (in units of $\hbar\omega_0$) as functions of $v_0/\Delta$ and $\omega_0 T$. The purple dash lines represent the curve $v_0T=(4n+2)\pi$ for (a) $n=0$ and (b) $n$ is a natural number. Other parameters are $\lambda=2$ and $N =10$.}
\label{fig3}
\end{figure}

During the charging process, not all parameters enable the QB to achieve full charging. For instance, in case $v_0/\Delta=10$, the QB can store only half as much energy as in fully charging. Therefore, it is crucial to determine the conditions that enables full charging of the QB. Neglecting the second term in Eq. (\ref{Et}) due to its being relatively smaller compared to the first term, the stored energy can be expressed as
\begin{equation}
\label{Et1}
E(t)=\frac{N\Delta}{2}\left[1+\cos \mu (t)\right].
\end{equation}
The maximum stored energy is given by
\begin{equation}
\label{Emax}
E_{\text{max}} = \begin{cases}
\dfrac{N\Delta}{2}\left[ {1 - \cos \left( {\dfrac{{{v_0}T}}{2}} \right)} \right], & 0 < {v_0}T <2\pi,\\
N\Delta, & {v_0}T \geq 2\pi,
\end{cases}
\end{equation}
then the parameter range is divided into regions of full charging and partial charging by the critical curve $v_0T=2\pi$, as depicted in Fig. \ref {fig3} (a).
Furthermore, we define the time corresponding to the attainment of the maximum stored energy as $t_{\text{max}}$. For the case of $0 < {v_0}T <2\pi$, $t_{max}=T$. When $v_0T \geq 2\pi$, $t_{max}$ satisfies the following equation:
\begin{equation}
\frac{v_0t_{max}}{2}-\frac{v_0T}{4\pi}\sin\left(\frac{2\pi t_{max}}{T}\right)=(2n+1)\pi,
\end{equation}
where $n$ is a natural number.

In our previous analysis, only the conditions for obtaining maximum stored energy are considered. However, the final stored energy is also an important factor for assessing battery performance. To do so, the stored energy at the end of charging is
\begin{equation}
\label{Etau}
E(\tau)=\frac{N\Delta}{2}\left[ {1-\cos \left( {\dfrac{{{v_0}T}}{2}} \right)} \right],
\end{equation}
which verifies the periodic evolution behaviors in Fig. \ref{fig3} (b). Here the purple dash lines are $v_0T=(4n+2)\pi$, the final stored energy reaches a maximum when the external field parameters meet this condition.

\section{The relationship among the stored energy, energy quantum fluctuations and diagonal entropy}\label{section4}
In this section we further discuss the relationship among the stored energy, energy quantum fluctuations, and diagonal entropy. The final stored energy is shown in Eq. (\ref{Etau}). Substituting Eq. (\ref{Et1}) into Eq. (\ref{sig}), the final quantum fluctuations as
\begin{figure}[htbp]
 \centering
 \includegraphics[width=0.485\textwidth]{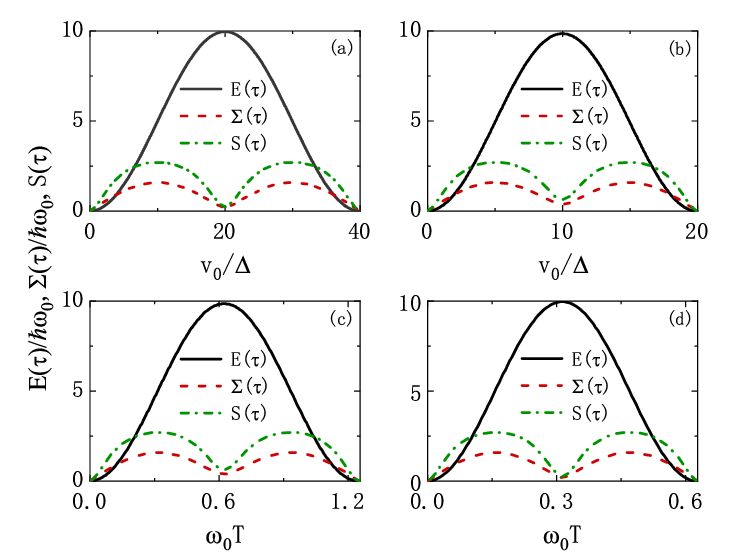}
 \caption{The final stored energy $E(\tau)$ (in units of $\hbar\omega_0$), final energy fluctuations $\Sigma(\tau)$ (in units of $\hbar\omega_0$) and final von Neumann entropy $S(\tau)$ as functions of (a) $\omega_0 T=0.1\pi$, (b) $\omega_0 T=0.2\pi$, (c) $v_0/\Delta=10$, and (d) $v_0/\Delta=20$, respectively. Black solid line: the final stored energy, red dash line: the final energy fluctuations, while green dash dot line: the final diagonal entropy. Other parameters are the same as in Fig. \ref{fig3}.}
\label{fig4}
\end{figure}

\begin{equation}
\Sigma(\tau) =\sqrt {\frac{N\Delta}{4}} \left| {\sin \left( {\frac{{{v_0}T}}{2}} \right)} \right|.
\end{equation}
For the diagonal entropy, since $B_2B_8B_9\cos \mu (t) - B_3B_8B_9\sin \mu (t)\ll 1$ in the Eq. (\ref{S}), the final diagonal entropy can be simplified to
\begin{equation}
S(\tau ) = {\log _2}\left( {\sqrt {(N-1){e^{\frac{\pi}{2}}}} } \right)\left| {\sin \left(\frac{{{v_0}T}}{2}\right)} \right|.
\end{equation}
The local maximum value of the final stored energy corresponds to the local minimum value of the final energy fluctuations and diagonal entropy, as depicted in Fig. \ref{fig4}, which is consistent with earlier results \cite{PhysRevLett.129.130602}.
Additionally, a partially synchronous relationship exists between the final energy quantum fluctuations and diagonal entropy. To further explore this relationship, Fig. \ref{fig5} presents the dependence of the final quantum fluctuations and diagonal entropy on $v_0/\Delta$ and $\omega_0T$.  The yellow dash lines represent the curve $v_0T=(4n+2)\pi, n=0,1,2,\cdots$. It is clearly seen that the their behavior remains consistent. This is further verified the conclusion that final energy quantum fluctuations and diagonal entropy exhibit their minimum values when the maximum stored energy exhibits its maximum value.
\begin{figure}[tbp]
 \centering
 \includegraphics[width=0.485\textwidth]{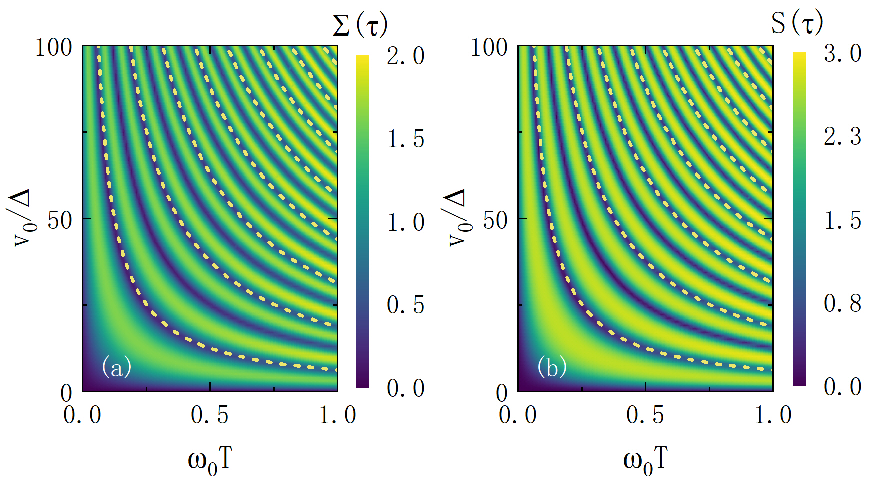}
\caption{Contour plots of QB's (a) final energy fluctuations and (b) final diagonal entropy as functions of $v_0/\Delta$ and $\omega_0 T$. The yellow dash lines represent the curve $v_0T=(4n+2)\pi, n=0,1,2,\cdots$. Other parameters are the same as in Fig. \ref{fig3}.}
\label{fig5}
\end{figure}
\section{The role of the atomic interactions and number of atoms}\label{section5}
\begin{figure}[tbp]
 \centering
 \includegraphics[width=0.485\textwidth]{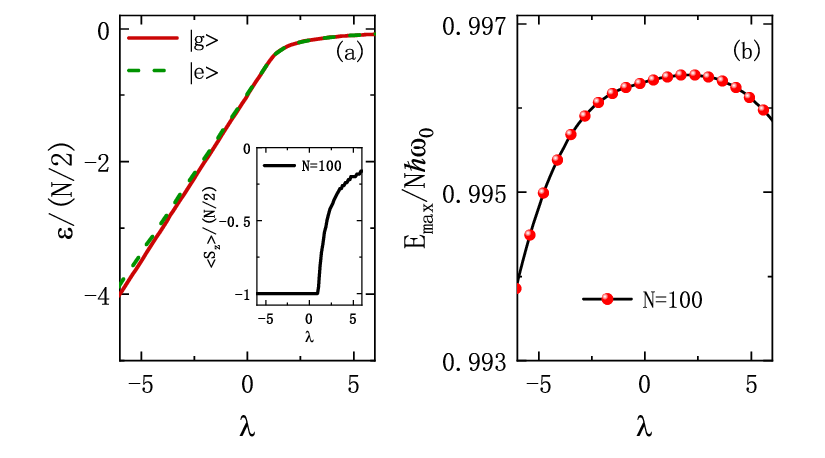}
 \caption{(a) The energy levels $\varepsilon/(N/2)$ for the ground state $|g\rangle$ (red solid line) and the first-excited state energy $|e\rangle$ (green dashed line) vs $\lambda$. The inset shows $\left\langle {\left. {{S_z}} \right\rangle } \right /(N/2)$ as a function of $\lambda$ for $N = 100$ TLSs. (b) The maximum stored energy $E_{max}$ (in unit of $N\hbar\omega_0$) vs $\lambda$ for $N=100$ TLSs. $v_0/\Delta=20$ and other parameters are the same as in Fig. \ref{fig2}.}
\label{fig6}
\end{figure}

Finally, we investigate the role of atomic interactions and number of TLSs on the charging performances of the QB.
Figure \ref{fig6} (a) displays the energy levels $\varepsilon/(N/2)$ of $N = 100$ TLSs in the ground state and the first-excited
state, which are almost degenerate for $\lambda>1$, whereas for $\lambda<1$, the energy levels are non-degenerate.
The inset in Fig. \ref {fig6} (a) shows the behavior of the order parameter $\left\langle {\left. {{S_z}} \right\rangle } \right /(N/2)$ for $N = 100$, exhibiting a quantum phase transition at $\lambda=1$. Furthermore, Fig. \ref{fig6} (b) presents the calculation result of the maximum stored energy $E_{max}/(N\hbar\omega_0)$ as a function of the atomic interactions strength. Near the critical point of the quantum phase transition, the maximum stored energy of the RZ quantum battery reaches its highest values.

\begin{figure}[htbp]
 \centering
 \includegraphics[width=0.485\textwidth]{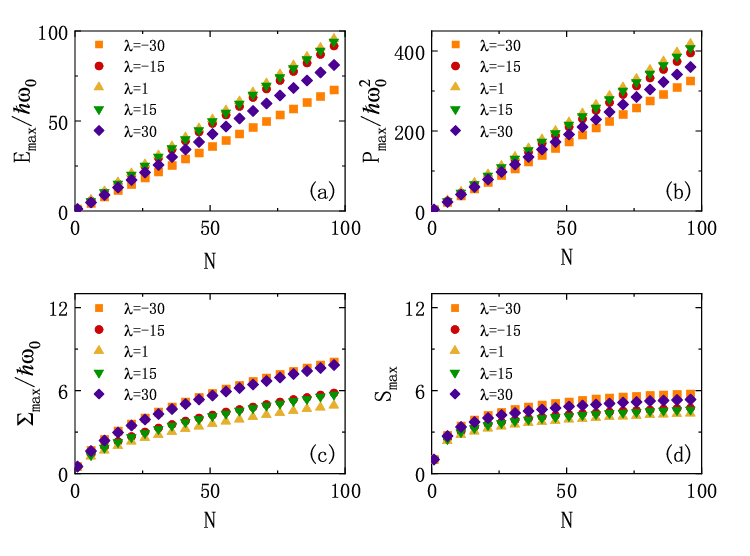}
 \caption{(a) Maximum stored energy $E_{max}$ (in units $\hbar\omega_0$), (b) maximum charging prower $P_{max}$ (in units $\hbar\omega_0^2$), (c) maximum quantum energy fluctuations $\Sigma_{max}$ (in units $\hbar\omega_0$) and (d) maximum diagonal entropy $S_{max}$ with the number $N$ $(N \in [1, 100])$ for different atomic interactions. Orange square, red circle, yellow up-triangle, green down-triangle and blue diamond represent $\lambda=-30,-15,1,15,30$, respectively. Other parameters are the same as in Fig. \ref{fig6}.}
\label{fig7}
\end{figure}

In Fig. \ref {fig7}, we show the QB's maximum stored energy, charging power, energy quantum fluctuations, and diagonal entropy as a function of the number $N$ of the TLSs for different atomic interactions. The maximum stored energy and charging power increase with the number of TLSs [see Fig. \ref {fig7} (a) and (b)]. Figure \ref {fig7} (c) illustrates that a stronger interatomic interaction results in higher energy uncertainty. Furthermore, the maximum diagonal entropy exhibits a logarithmic growth trend as the number of atoms increases. 
 Increasing the number of TLSs enhances the quantum coherence effect, which can further promote the transfer of energy between the battery and the charger.
\section{Conclusions}\label{section6}
In conclusion, we have constructed the RZ quantum battery constituted of $N$ TLSs and charged by both the driving field and atomic interactions.
By employing a gauge transformation, we have obtained an analytical solution for the QB, which demonstrated good agreement with numerical results for cases with a small external field scanning period.
The condition for the full charging of the QB has been determined to be $v_0T\geq2\pi$, and the maximum final stored energy is achieved when $v_0T=(4n+2)\pi$.
Furthermore, the local maximum value of the final stored energy corresponds to the local minimum value of the final uncertainty and entanglement. We have also analyzed the effect of atomic interactions and the number of TLSs on the QB's stored energy, charging power, energy quantum fluctuations and diagonal entropy. Our analysis revealed that the maximum stored energy reaches its highest values in the proximity of the quantum phase transition point $\lambda=1$. Additionally, the maximum stored energy and charging power increase with the number of TLSs. Increasing the number of TLSs and atomic interactions will enhance the uncertainty of energy, but it also further promotes the transfer of energy between the battery and the charger.

We can also further consider how to enhance the performance of the QB through the optimal control \cite{PhysRevA.107.032218, PhysRevE.90.012119,PhysRevA.108.062425}. Recently, many efforts have been devoted to actual implementations of the QB, including the Dicke QB \cite{Superabsorption2022}, the star-topology nuclear magnetic resonance spin systems QB \cite{PhysRevA.106.042601}, the solid-state qubit QB \cite{wenniger2022coherence,batteries8050043}, the transmon qutrit QB \cite{Hu_2022}, the transmon qubit-resonator QB \cite{PhysRevA.107.023725}, the xmon qutrit QB \cite{Zheng_2022}, and the XXZ Heisenberg QB \cite{PhysRevA.107.L030201}. In addition, experimental efforts have been devoted to quantum simulations of an array of TLSs, such as cold atoms \cite{PhysRevA.82.053841}, trapped ions \cite{PhysRevLett.74.4091, Bruzewicz2019, Georgescu2020, PhysRevX.8.021027}, and quantum dots in semiconductors \cite{A.2011}, which could be considered as QB.  When charging resources such as Raman laser beams are used to such TLSs, charging of the QB could be realistically implemented  \cite{PhysRevLett.120.117702, PhysRevE.99.052106, chen2020}.
This study aims to providing a more efficient two-level QB theoretical background for future experimental implementations.
\section*{Acknowledgments}
We thank S. C. Li, H. Cao, D. L. Yang, and Z. T. Zhang for helpful discussions. The work is supported by the National Natural Science Foundation of China (Grant No. 12075193).
\appendix
\section{}\label{app1}
The coefficients before the angular momentum operator after the second unitary transformation are as follows:
\begin{equation}
	\begin{aligned}
	A_1(t)&=v_0 \cos\nu(t) \sin^2 \left( \dfrac{\pi t}{T} \right)
    +\dot{\mu}(t)\cos\nu(t)\\&-\Delta\cos\mu(t)\sin\nu(t),\\
	A_2(t)&=\dot{\nu}(t)+\Delta\sin\mu(t),\\
	A_3(t)&=-v_0\sin\nu(t)\sin^2 \left( \dfrac{\pi t}{T} \right)
    -\dot{\mu}(t)\sin\nu(t)\\&+\Delta\cos\mu(t)\cos\nu(t),\\	
    A_4(t)&=A_5(t)=-\frac{2\lambda }{N}\sin\mu(t)\cos\mu(t)\sin\nu(t),\\
	A_6(t)&=A_7(t)=-\frac{2\lambda }{N}\sin\nu(t)\cos\nu(t)\cos^2\mu(t),\\
	A_8(t)&=A_9(t)=\frac{2\lambda }{N}\sin\mu(t)\cos\mu(t)\cos\nu(t),\\
    A_{10}(t)&=\frac{2\lambda }{N}\cos^2\mu(t)\sin^2\nu(t),\\
    A_{11}(t)&=\frac{2\lambda }{N}\sin^2\mu(t),\\	
	A_{12}(t)&=\frac{2\lambda }{N}\cos^2\mu(t)\cos^2\nu(t).	
 	\end{aligned}
\end{equation}
\section{}\label{app2}
We replace the composite trigonometric function with the following Bessel equation $J_n$ in the integration process of Eq. (\ref{U3}),
and neglecting the term $n\geq2$ in $J_n$ as the approximation,
\begin{equation}
\begin{aligned}
\cos \left( {z\sin {\phi _t}} \right) =& {J_0}(z) + \sum\nolimits_{k = 1}^\infty  {2{J_{2k}}(z)\cos \left( {2k{\phi_t}} \right)}, \\
\sin \left( {z\sin {\phi _t}} \right) =& \sum\nolimits_{k = 0}^\infty  {2{J_{2k + 1}}(z)\sin \left[ {\left( {2k + 1} \right){\phi _t}} \right]}.
\end{aligned}
\end{equation}
The expression after the integration is
\begin{equation}
\begin{aligned}
{B_1}& = {B_4} = {B_5} = {B_6} = {B_7} = {B_{10}} = 0,\\
B_2 &= 4\Delta\left[ {\frac{{{J_0}\left( { - \frac{{{v_0}T}}{{4\pi }}} \right)}}{{{v_0}}} + \frac{{8\pi T{J_1}\left( { - \frac{{{v_0}T}}{{4\pi }}} \right)}}{{16{\pi ^2} - {v_0}^2{T^2}}}} \right]{\sin ^2}\left( {\frac{{{v_0}T}}{4}} \right),\\
{B_3} &=  - 2{\Delta}\left[ {\frac{{{J_0}\left( { - \frac{{{v_0}T}}{{4\pi }}} \right)}}{{{v_0}}} + \frac{{8\pi T{J_1}\left( { - \frac{{{v_0}T}}{{4\pi }}} \right)}}{{16{\pi ^2} - {v_0}^2{T^2}}}} \right]\sin \left( {\frac{{{v_0}T}}{2}} \right),\\
{B_8} &= \frac{\lambda }{N}\left[ {\frac{{{J_0}\left( { - \frac{{{v_0}T}}{{2\pi }}} \right)}}{{{v_0}}} + \frac{{4\pi T{J_1}\left( { - \frac{{{v_0}T}}{{2\pi }}} \right)}}{{4{\pi ^2} - {v_0}^2{T^2}}}} \right]\left[ {1 - \cos \left( {{v_0}T} \right)} \right], \\
{B_9} &= \frac{\lambda }{N}\left[ {\frac{{{J_0}\left( { - \frac{{{v_0}T}}{{2\pi }}} \right)}}{{{v_0}}} + \frac{{4\pi T{J_1}\left( { - \frac{{{v_0}T}}{{2\pi }}} \right)}}{{4{\pi ^2} - {v_0}^2{T^2}}}} \right]\left[ {1 - \cos \left( {{v_0}T} \right)} \right],\\
B_{11} &= \frac{\lambda }{N}\left[{{ T-\left(\frac{{J_0\left( { - \frac{{{v_0}T}}{{2\pi }}} \right)}}{{{v_0}}} - \frac{{4\pi T{J_1}\left( { - \frac{{{v_0}T}}{{2\pi }}} \right)}}{{{v_0}^2{T^2} - 4{\pi ^2}}}\right)\sin \left( {{v_0}T} \right)}} \right],\\
B_{12} &= \frac{\lambda }{N}\left[ {{T+\left(\frac{{J_0\left( { - \frac{{{v_0}T}}{{2\pi }}} \right)}}{{{v_0}}} - \frac{{4\pi T{J_1}\left( { - \frac{{{v_0}T}}{{2\pi }}} \right)}}{{{v_0}^2{T^2} - 4{\pi ^2}}}\right)\sin \left( {{v_0}T} \right)}} \right].
\end{aligned}
\end{equation}
\section{}\label{app3}
In the process of solving the expression of energy fluctuations, we apply $U_1(t)$ and $U_2(t)$ to $J_z^2$. After the first two unitary evolutions, we obtain the following results:
\begin{equation}
\begin{aligned}
&{U_1}(t)(J_z^2)U_1^\dag(t)= \frac{1}{2}\sin2\mu(t)({J_z}{J_y} + {J_y}{J_z})\\
&+ \frac{1}{2}\left[ {1 + \cos 2\mu(t) } \right]J_z^2 + \frac{1}{2}\left[ {1 - \cos 2\mu(t) } \right]J_y^2,
\end{aligned}
\end{equation}
\begin{equation}
\begin{aligned}
&{U_2(t)}{U_1(t)}(J_z^2)U_1^\dag(t) U_2^\dag(t)= {C_1(t)}{J_x}{J_y} + {C_2(t)}{J_y}{J_x}\\
&+ {C_3(t)}{J_x}{J_z}+{C_4(t)}{J_z}{J_x} + {C_5(t)}{J_y}{J_z}+ {C_6(t)}{J_z}{J_y}\\
&+{C_7(t)}J_x^2 + {C_8(t)}J_y^2 + {C_9(t)}J_z^2,
\end{aligned}
\end{equation}
where
\begin{equation}
\begin{aligned}
&C_1(t)=C_2(t)=-\sin\mu(t)\cos\mu(t)\sin\nu(t),\\
&C_3(t)=C_4(t)=-\sin\nu(t)\cos\nu(t)\cos^2\mu(t),\\
&C_5(t)=C_6(t)=\sin\mu(t)\cos\mu(t)\cos\nu(t),\\
&C_7(t)=\cos^2\mu(t)\sin^2\nu(t),\\
&C_8(t)=\sin^2\mu(t),\\
&C_9(t)=\cos^2\mu(t)\cos^2\nu(t).
\end{aligned}
\end{equation}
\bibliography{reference}

\begin{thebibliography}{108}%
\makeatletter
\providecommand \@ifxundefined [1]{%
 \@ifx{#1\undefined}
}%
\providecommand \@ifnum [1]{%
 \ifnum #1\expandafter \@firstoftwo
 \else \expandafter \@secondoftwo
 \fi
}%
\providecommand \@ifx [1]{%
 \ifx #1\expandafter \@firstoftwo
 \else \expandafter \@secondoftwo
 \fi
}%
\providecommand \natexlab [1]{#1}%
\providecommand \enquote  [1]{``#1''}%
\providecommand \bibnamefont  [1]{#1}%
\providecommand \bibfnamefont [1]{#1}%
\providecommand \citenamefont [1]{#1}%
\providecommand \href@noop [0]{\@secondoftwo}%
\providecommand \href [0]{\begingroup \@sanitize@url \@href}%
\providecommand \@href[1]{\@@startlink{#1}\@@href}%
\providecommand \@@href[1]{\endgroup#1\@@endlink}%
\providecommand \@sanitize@url [0]{\catcode `\\12\catcode `\$12\catcode
  `\&12\catcode `\#12\catcode `\^12\catcode `\_12\catcode `\%12\relax}%
\providecommand \@@startlink[1]{}%
\providecommand \@@endlink[0]{}%
\providecommand \url  [0]{\begingroup\@sanitize@url \@url }%
\providecommand \@url [1]{\endgroup\@href {#1}{\urlprefix }}%
\providecommand \urlprefix  [0]{URL }%
\providecommand \Eprint [0]{\href }%
\providecommand \doibase [0]{http://dx.doi.org/}%
\providecommand \selectlanguage [0]{\@gobble}%
\providecommand \bibinfo  [0]{\@secondoftwo}%
\providecommand \bibfield  [0]{\@secondoftwo}%
\providecommand \translation [1]{[#1]}%
\providecommand \BibitemOpen [0]{}%
\providecommand \bibitemStop [0]{}%
\providecommand \bibitemNoStop [0]{.\EOS\space}%
\providecommand \EOS [0]{\spacefactor3000\relax}%
\providecommand \BibitemShut  [1]{\csname bibitem#1\endcsname}%
\let\auto@bib@innerbib\@empty
\bibitem [{\citenamefont {Juli\`a-Farr\'e}\ \emph {et~al.}(2020)\citenamefont
  {Juli\`a-Farr\'e}, \citenamefont {Salamon}, \citenamefont {Riera},
  \citenamefont {Bera},\ and\ \citenamefont
  {Lewenstein}}]{PhysRevResearch.2.023113}%
  \BibitemOpen
  \bibfield  {author} {\bibinfo {author} {\bibfnamefont {S.}~\bibnamefont
  {Juli\`a-Farr\'e}}, \bibinfo {author} {\bibfnamefont {T.}~\bibnamefont
  {Salamon}}, \bibinfo {author} {\bibfnamefont {A.}~\bibnamefont {Riera}},
  \bibinfo {author} {\bibfnamefont {M.~N.}\ \bibnamefont {Bera}}, \ and\
  \bibinfo {author} {\bibfnamefont {M.}~\bibnamefont {Lewenstein}},\ }\href
  {\doibase 10.1103/PhysRevResearch.2.023113} {\bibfield  {journal} {\bibinfo
  {journal} {Phys. Rev. Res.}\ }\textbf {\bibinfo {volume} {2}},\ \bibinfo
  {pages} {023113} (\bibinfo {year} {2020})}\BibitemShut {NoStop}%
\bibitem [{\citenamefont {Carrega}\ \emph {et~al.}(2019)\citenamefont
  {Carrega}, \citenamefont {Sassetti},\ and\ \citenamefont
  {Weiss}}]{PhysRevA.99.062111}%
  \BibitemOpen
  \bibfield  {author} {\bibinfo {author} {\bibfnamefont {M.}~\bibnamefont
  {Carrega}}, \bibinfo {author} {\bibfnamefont {M.}~\bibnamefont {Sassetti}}, \
  and\ \bibinfo {author} {\bibfnamefont {U.}~\bibnamefont {Weiss}},\ }\href
  {\doibase 10.1103/PhysRevA.99.062111} {\bibfield  {journal} {\bibinfo
  {journal} {Phys. Rev. A}\ }\textbf {\bibinfo {volume} {99}},\ \bibinfo
  {pages} {062111} (\bibinfo {year} {2019})}\BibitemShut {NoStop}%
\bibitem [{\citenamefont {Francica}\ \emph {et~al.}(2020)\citenamefont
  {Francica}, \citenamefont {Binder}, \citenamefont {Guarnieri}, \citenamefont
  {Mitchison}, \citenamefont {Goold},\ and\ \citenamefont
  {Plastina}}]{PhysRevLett.125.180603}%
  \BibitemOpen
  \bibfield  {author} {\bibinfo {author} {\bibfnamefont {G.}~\bibnamefont
  {Francica}}, \bibinfo {author} {\bibfnamefont {F.~C.}\ \bibnamefont
  {Binder}}, \bibinfo {author} {\bibfnamefont {G.}~\bibnamefont {Guarnieri}},
  \bibinfo {author} {\bibfnamefont {M.~T.}\ \bibnamefont {Mitchison}}, \bibinfo
  {author} {\bibfnamefont {J.}~\bibnamefont {Goold}}, \ and\ \bibinfo {author}
  {\bibfnamefont {F.}~\bibnamefont {Plastina}},\ }\href {\doibase
  10.1103/PhysRevLett.125.180603} {\bibfield  {journal} {\bibinfo  {journal}
  {Phys. Rev. Lett.}\ }\textbf {\bibinfo {volume} {125}},\ \bibinfo {pages}
  {180603} (\bibinfo {year} {2020})}\BibitemShut {NoStop}%
\bibitem [{\citenamefont {Mukherjee}\ and\ \citenamefont
  {Divakaran}(2021)}]{Mukherjee_2021}%
  \BibitemOpen
  \bibfield  {author} {\bibinfo {author} {\bibfnamefont {V.}~\bibnamefont
  {Mukherjee}}\ and\ \bibinfo {author} {\bibfnamefont {U.}~\bibnamefont
  {Divakaran}},\ }\href {\doibase 10.1088/1361-648X/ac1b60} {\bibfield
  {journal} {\bibinfo  {journal} {Journal of Physics: Condensed Matter}\
  }\textbf {\bibinfo {volume} {33}},\ \bibinfo {pages} {454001} (\bibinfo
  {year} {2021})}\BibitemShut {NoStop}%
\bibitem [{\citenamefont {Watanabe}\ \emph {et~al.}(2020)\citenamefont
  {Watanabe}, \citenamefont {Venkatesh}, \citenamefont {Talkner}, \citenamefont
  {Hwang},\ and\ \citenamefont {del Campo}}]{PhysRevLett.124.210603}%
  \BibitemOpen
  \bibfield  {author} {\bibinfo {author} {\bibfnamefont {G.}~\bibnamefont
  {Watanabe}}, \bibinfo {author} {\bibfnamefont {B.~P.}\ \bibnamefont
  {Venkatesh}}, \bibinfo {author} {\bibfnamefont {P.}~\bibnamefont {Talkner}},
  \bibinfo {author} {\bibfnamefont {M.-J.}\ \bibnamefont {Hwang}}, \ and\
  \bibinfo {author} {\bibfnamefont {A.}~\bibnamefont {del Campo}},\ }\href
  {\doibase 10.1103/PhysRevLett.124.210603} {\bibfield  {journal} {\bibinfo
  {journal} {Phys. Rev. Lett.}\ }\textbf {\bibinfo {volume} {124}},\ \bibinfo
  {pages} {210603} (\bibinfo {year} {2020})}\BibitemShut {NoStop}%
\bibitem [{\citenamefont {Revathy}\ \emph {et~al.}(2020)\citenamefont
  {Revathy}, \citenamefont {Mukherjee}, \citenamefont {Divakaran},\ and\
  \citenamefont {del Campo}}]{PhysRevResearch.2.043247}%
  \BibitemOpen
  \bibfield  {author} {\bibinfo {author} {\bibfnamefont {B.~S.}\ \bibnamefont
  {Revathy}}, \bibinfo {author} {\bibfnamefont {V.}~\bibnamefont {Mukherjee}},
  \bibinfo {author} {\bibfnamefont {U.}~\bibnamefont {Divakaran}}, \ and\
  \bibinfo {author} {\bibfnamefont {A.}~\bibnamefont {del Campo}},\ }\href
  {\doibase 10.1103/PhysRevResearch.2.043247} {\bibfield  {journal} {\bibinfo
  {journal} {Phys. Rev. Res.}\ }\textbf {\bibinfo {volume} {2}},\ \bibinfo
  {pages} {043247} (\bibinfo {year} {2020})}\BibitemShut {NoStop}%
\bibitem [{\citenamefont {Stefanatos}(2014)}]{PhysRevE.90.012119}%
  \BibitemOpen
  \bibfield  {author} {\bibinfo {author} {\bibfnamefont {D.}~\bibnamefont
  {Stefanatos}},\ }\href {\doibase 10.1103/PhysRevE.90.012119} {\bibfield
  {journal} {\bibinfo  {journal} {Phys. Rev. E}\ }\textbf {\bibinfo {volume}
  {90}},\ \bibinfo {pages} {012119} (\bibinfo {year} {2014})}\BibitemShut
  {NoStop}%
\bibitem [{\citenamefont {Millen}\ and\ \citenamefont
  {Xuereb}(2016)}]{Millen_2016}%
  \BibitemOpen
  \bibfield  {author} {\bibinfo {author} {\bibfnamefont {J.}~\bibnamefont
  {Millen}}\ and\ \bibinfo {author} {\bibfnamefont {A.}~\bibnamefont
  {Xuereb}},\ }\href {\doibase 10.1088/2058-7058/29/1/30} {\bibfield  {journal}
  {\bibinfo  {journal} {Phys. World}\ }\textbf {\bibinfo {volume} {29}},\
  \bibinfo {pages} {23} (\bibinfo {year} {2016})}\BibitemShut {NoStop}%
\bibitem [{\citenamefont {Pirmoradian}\ and\ \citenamefont
  {M\o{}lmer}(2019)}]{PhysRevA.100.043833}%
  \BibitemOpen
  \bibfield  {author} {\bibinfo {author} {\bibfnamefont {F.}~\bibnamefont
  {Pirmoradian}}\ and\ \bibinfo {author} {\bibfnamefont {K.}~\bibnamefont
  {M\o{}lmer}},\ }\href {\doibase 10.1103/PhysRevA.100.043833} {\bibfield
  {journal} {\bibinfo  {journal} {Phys. Rev. A}\ }\textbf {\bibinfo {volume}
  {100}},\ \bibinfo {pages} {043833} (\bibinfo {year} {2019})}\BibitemShut
  {NoStop}%
\bibitem [{\citenamefont {Centrone}\ \emph {et~al.}()\citenamefont {Centrone},
  \citenamefont {Mancino},\ and\ \citenamefont
  {Paternostro}}]{centrone2021charging}%
  \BibitemOpen
  \bibfield  {author} {\bibinfo {author} {\bibfnamefont {F.}~\bibnamefont
  {Centrone}}, \bibinfo {author} {\bibfnamefont {L.}~\bibnamefont {Mancino}}, \
  and\ \bibinfo {author} {\bibfnamefont {M.}~\bibnamefont {Paternostro}},\
  }\href@noop {} {}\Eprint {http://arxiv.org/abs/arXiv: 2106. 07899} {arXiv:
  2106. 07899} \BibitemShut {NoStop}%
\bibitem [{\citenamefont {Zhao}\ \emph {et~al.}(2022)\citenamefont {Zhao},
  \citenamefont {Dou},\ and\ \citenamefont {Zhao}}]{PhysRevResearch.4.013172}%
  \BibitemOpen
  \bibfield  {author} {\bibinfo {author} {\bibfnamefont {F.}~\bibnamefont
  {Zhao}}, \bibinfo {author} {\bibfnamefont {F.-Q.}\ \bibnamefont {Dou}}, \
  and\ \bibinfo {author} {\bibfnamefont {Q.}~\bibnamefont {Zhao}},\ }\href
  {\doibase 10.1103/PhysRevResearch.4.013172} {\bibfield  {journal} {\bibinfo
  {journal} {Phys. Rev. Res.}\ }\textbf {\bibinfo {volume} {4}},\ \bibinfo
  {pages} {013172} (\bibinfo {year} {2022})}\BibitemShut {NoStop}%
\bibitem [{\citenamefont {Alicki}\ and\ \citenamefont
  {Fannes}(2013)}]{PhysRevE.87.042123}%
  \BibitemOpen
  \bibfield  {author} {\bibinfo {author} {\bibfnamefont {R.}~\bibnamefont
  {Alicki}}\ and\ \bibinfo {author} {\bibfnamefont {M.}~\bibnamefont
  {Fannes}},\ }\href {\doibase 10.1103/PhysRevE.87.042123} {\bibfield
  {journal} {\bibinfo  {journal} {Phys. Rev. E}\ }\textbf {\bibinfo {volume}
  {87}},\ \bibinfo {pages} {042123} (\bibinfo {year} {2013})}\BibitemShut
  {NoStop}%
\bibitem [{\citenamefont {Campaioli}\ \emph {et~al.}(2023)\citenamefont
  {Campaioli}, \citenamefont {Gherardini}, \citenamefont {Quach}, \citenamefont
  {Polini},\ and\ \citenamefont {Andolina}}]{campaioli2023colloquium}%
  \BibitemOpen
  \bibfield  {author} {\bibinfo {author} {\bibfnamefont {F.}~\bibnamefont
  {Campaioli}}, \bibinfo {author} {\bibfnamefont {S.}~\bibnamefont
  {Gherardini}}, \bibinfo {author} {\bibfnamefont {J.~Q.}\ \bibnamefont
  {Quach}}, \bibinfo {author} {\bibfnamefont {M.}~\bibnamefont {Polini}}, \
  and\ \bibinfo {author} {\bibfnamefont {G.~M.}\ \bibnamefont {Andolina}},\
  }\href@noop {} {\enquote {\bibinfo {title} {Colloquium: Quantum batteries},}\
  } (\bibinfo {year} {2023}),\ \Eprint {http://arxiv.org/abs/2308.02277}
  {arXiv:2308.02277 [quant-ph]} \BibitemShut {NoStop}%
\bibitem [{\citenamefont {Zhao}\ \emph {et~al.}(2021)\citenamefont {Zhao},
  \citenamefont {Dou},\ and\ \citenamefont {Zhao}}]{PhysRevA.103.033715}%
  \BibitemOpen
  \bibfield  {author} {\bibinfo {author} {\bibfnamefont {F.}~\bibnamefont
  {Zhao}}, \bibinfo {author} {\bibfnamefont {F.-Q.}\ \bibnamefont {Dou}}, \
  and\ \bibinfo {author} {\bibfnamefont {Q.}~\bibnamefont {Zhao}},\ }\href
  {\doibase 10.1103/PhysRevA.103.033715} {\bibfield  {journal} {\bibinfo
  {journal} {Phys. Rev. A}\ }\textbf {\bibinfo {volume} {103}},\ \bibinfo
  {pages} {033715} (\bibinfo {year} {2021})}\BibitemShut {NoStop}%
\bibitem [{\citenamefont {Dou}\ \emph {et~al.}(2020)\citenamefont {Dou},
  \citenamefont {Wang},\ and\ \citenamefont {Sun}}]{Dou_2020}%
  \BibitemOpen
  \bibfield  {author} {\bibinfo {author} {\bibfnamefont {F.-Q.}\ \bibnamefont
  {Dou}}, \bibinfo {author} {\bibfnamefont {Y.-J.}\ \bibnamefont {Wang}}, \
  and\ \bibinfo {author} {\bibfnamefont {J.-A.}\ \bibnamefont {Sun}},\ }\href
  {\doibase 10.1209/0295-5075/131/43001} {\bibfield  {journal} {\bibinfo
  {journal} {Europhys. Lett.}\ }\textbf {\bibinfo {volume} {131}},\ \bibinfo
  {pages} {43001} (\bibinfo {year} {2020})}\BibitemShut {NoStop}%
\bibitem [{\citenamefont {Caravelli}\ \emph {et~al.}(2020)\citenamefont
  {Caravelli}, \citenamefont {Coulter-De~Wit}, \citenamefont
  {Garc\'{\i}a-Pintos},\ and\ \citenamefont
  {Hamma}}]{PhysRevResearch.2.023095}%
  \BibitemOpen
  \bibfield  {author} {\bibinfo {author} {\bibfnamefont {F.}~\bibnamefont
  {Caravelli}}, \bibinfo {author} {\bibfnamefont {G.}~\bibnamefont
  {Coulter-De~Wit}}, \bibinfo {author} {\bibfnamefont {L.~P.}\ \bibnamefont
  {Garc\'{\i}a-Pintos}}, \ and\ \bibinfo {author} {\bibfnamefont
  {A.}~\bibnamefont {Hamma}},\ }\href {\doibase
  10.1103/PhysRevResearch.2.023095} {\bibfield  {journal} {\bibinfo  {journal}
  {Phys. Rev. Res.}\ }\textbf {\bibinfo {volume} {2}},\ \bibinfo {pages}
  {023095} (\bibinfo {year} {2020})}\BibitemShut {NoStop}%
\bibitem [{\citenamefont {Quach}\ and\ \citenamefont
  {Munro}(2020)}]{PhysRevApplied.14.024092}%
  \BibitemOpen
  \bibfield  {author} {\bibinfo {author} {\bibfnamefont {J.~Q.}\ \bibnamefont
  {Quach}}\ and\ \bibinfo {author} {\bibfnamefont {W.~J.}\ \bibnamefont
  {Munro}},\ }\href {\doibase 10.1103/PhysRevApplied.14.024092} {\bibfield
  {journal} {\bibinfo  {journal} {Phys. Rev. Applied}\ }\textbf {\bibinfo
  {volume} {14}},\ \bibinfo {pages} {024092} (\bibinfo {year}
  {2020})}\BibitemShut {NoStop}%
\bibitem [{\citenamefont {Andolina}\ \emph {et~al.}(2018)\citenamefont
  {Andolina}, \citenamefont {Farina}, \citenamefont {Mari}, \citenamefont
  {Pellegrini}, \citenamefont {Giovannetti},\ and\ \citenamefont
  {Polini}}]{PhysRevB.98.205423}%
  \BibitemOpen
  \bibfield  {author} {\bibinfo {author} {\bibfnamefont {G.~M.}\ \bibnamefont
  {Andolina}}, \bibinfo {author} {\bibfnamefont {D.}~\bibnamefont {Farina}},
  \bibinfo {author} {\bibfnamefont {A.}~\bibnamefont {Mari}}, \bibinfo {author}
  {\bibfnamefont {V.}~\bibnamefont {Pellegrini}}, \bibinfo {author}
  {\bibfnamefont {V.}~\bibnamefont {Giovannetti}}, \ and\ \bibinfo {author}
  {\bibfnamefont {M.}~\bibnamefont {Polini}},\ }\href {\doibase
  10.1103/PhysRevB.98.205423} {\bibfield  {journal} {\bibinfo  {journal} {Phys.
  Rev. B}\ }\textbf {\bibinfo {volume} {98}},\ \bibinfo {pages} {205423}
  (\bibinfo {year} {2018})}\BibitemShut {NoStop}%
\bibitem [{\citenamefont {Zhang}\ \emph {et~al.}(2019)\citenamefont {Zhang},
  \citenamefont {Yang}, \citenamefont {Fu},\ and\ \citenamefont
  {Wang}}]{PhysRevE.99.052106}%
  \BibitemOpen
  \bibfield  {author} {\bibinfo {author} {\bibfnamefont {Y.-Y.}\ \bibnamefont
  {Zhang}}, \bibinfo {author} {\bibfnamefont {T.-R.}\ \bibnamefont {Yang}},
  \bibinfo {author} {\bibfnamefont {L.}~\bibnamefont {Fu}}, \ and\ \bibinfo
  {author} {\bibfnamefont {X.}~\bibnamefont {Wang}},\ }\href {\doibase
  10.1103/PhysRevE.99.052106} {\bibfield  {journal} {\bibinfo  {journal} {Phys.
  Rev. E}\ }\textbf {\bibinfo {volume} {99}},\ \bibinfo {pages} {052106}
  (\bibinfo {year} {2019})}\BibitemShut {NoStop}%
\bibitem [{\citenamefont {Crescente}\ \emph
  {et~al.}(2020{\natexlab{a}})\citenamefont {Crescente}, \citenamefont
  {Carrega}, \citenamefont {Sassetti},\ and\ \citenamefont
  {Ferraro}}]{Crescente_2020}%
  \BibitemOpen
  \bibfield  {author} {\bibinfo {author} {\bibfnamefont {A.}~\bibnamefont
  {Crescente}}, \bibinfo {author} {\bibfnamefont {M.}~\bibnamefont {Carrega}},
  \bibinfo {author} {\bibfnamefont {M.}~\bibnamefont {Sassetti}}, \ and\
  \bibinfo {author} {\bibfnamefont {D.}~\bibnamefont {Ferraro}},\ }\href
  {\doibase 10.1088/1367-2630/ab91fc} {\bibfield  {journal} {\bibinfo
  {journal} {New J. Phys.}\ }\textbf {\bibinfo {volume} {22}},\ \bibinfo
  {pages} {063057} (\bibinfo {year} {2020}{\natexlab{a}})}\BibitemShut
  {NoStop}%
\bibitem [{\citenamefont {Dou}\ \emph {et~al.}()\citenamefont {Dou},
  \citenamefont {Wang},\ and\ \citenamefont {Sun}}]{Dou2022charging}%
  \BibitemOpen
  \bibfield  {author} {\bibinfo {author} {\bibfnamefont {F.-Q.}\ \bibnamefont
  {Dou}}, \bibinfo {author} {\bibfnamefont {Y.-J.}\ \bibnamefont {Wang}}, \
  and\ \bibinfo {author} {\bibfnamefont {J.-A.}\ \bibnamefont {Sun}},\
  }\href@noop {} {}\Eprint {http://arxiv.org/abs/arXiv: 2208. 04831} {arXiv:
  2208. 04831} \BibitemShut {NoStop}%
\bibitem [{\citenamefont {Delmonte}\ \emph {et~al.}(2021)\citenamefont
  {Delmonte}, \citenamefont {Crescente}, \citenamefont {Carrega}, \citenamefont
  {Ferraro},\ and\ \citenamefont {Sassetti}}]{e23050612}%
  \BibitemOpen
  \bibfield  {author} {\bibinfo {author} {\bibfnamefont {A.}~\bibnamefont
  {Delmonte}}, \bibinfo {author} {\bibfnamefont {A.}~\bibnamefont {Crescente}},
  \bibinfo {author} {\bibfnamefont {M.}~\bibnamefont {Carrega}}, \bibinfo
  {author} {\bibfnamefont {D.}~\bibnamefont {Ferraro}}, \ and\ \bibinfo
  {author} {\bibfnamefont {M.}~\bibnamefont {Sassetti}},\ }\href
  {https://www.mdpi.com/1099-4300/23/5/612} {\bibfield  {journal} {\bibinfo
  {journal} {Entropy}\ }\textbf {\bibinfo {volume} {23}},\ \bibinfo {pages}
  {612} (\bibinfo {year} {2021})}\BibitemShut {NoStop}%
\bibitem [{\citenamefont {Mitchison}\ \emph {et~al.}(2021)\citenamefont
  {Mitchison}, \citenamefont {Goold},\ and\ \citenamefont
  {Prior}}]{Mitchison2021chargingquantum}%
  \BibitemOpen
  \bibfield  {author} {\bibinfo {author} {\bibfnamefont {M.~T.}\ \bibnamefont
  {Mitchison}}, \bibinfo {author} {\bibfnamefont {J.}~\bibnamefont {Goold}}, \
  and\ \bibinfo {author} {\bibfnamefont {J.}~\bibnamefont {Prior}},\ }\href
  {\doibase 10.22331/q-2021-07-13-500} {\bibfield  {journal} {\bibinfo
  {journal} {{Quantum}}\ }\textbf {\bibinfo {volume} {5}},\ \bibinfo {pages}
  {500} (\bibinfo {year} {2021})}\BibitemShut {NoStop}%
\bibitem [{\citenamefont {Andolina}\ \emph
  {et~al.}(2019{\natexlab{a}})\citenamefont {Andolina}, \citenamefont {Keck},
  \citenamefont {Mari}, \citenamefont {Campisi}, \citenamefont {Giovannetti},\
  and\ \citenamefont {Polini}}]{PhysRevLett.122.047702}%
  \BibitemOpen
  \bibfield  {author} {\bibinfo {author} {\bibfnamefont {G.~M.}\ \bibnamefont
  {Andolina}}, \bibinfo {author} {\bibfnamefont {M.}~\bibnamefont {Keck}},
  \bibinfo {author} {\bibfnamefont {A.}~\bibnamefont {Mari}}, \bibinfo {author}
  {\bibfnamefont {M.}~\bibnamefont {Campisi}}, \bibinfo {author} {\bibfnamefont
  {V.}~\bibnamefont {Giovannetti}}, \ and\ \bibinfo {author} {\bibfnamefont
  {M.}~\bibnamefont {Polini}},\ }\href {\doibase
  10.1103/PhysRevLett.122.047702} {\bibfield  {journal} {\bibinfo  {journal}
  {Phys. Rev. Lett.}\ }\textbf {\bibinfo {volume} {122}},\ \bibinfo {pages}
  {047702} (\bibinfo {year} {2019}{\natexlab{a}})}\BibitemShut {NoStop}%
\bibitem [{\citenamefont {Garc\'{\i}a-Pintos}\ \emph
  {et~al.}(2020)\citenamefont {Garc\'{\i}a-Pintos}, \citenamefont {Hamma},\
  and\ \citenamefont {del Campo}}]{PhysRevLett.125.040601}%
  \BibitemOpen
  \bibfield  {author} {\bibinfo {author} {\bibfnamefont {L.~P.}\ \bibnamefont
  {Garc\'{\i}a-Pintos}}, \bibinfo {author} {\bibfnamefont {A.}~\bibnamefont
  {Hamma}}, \ and\ \bibinfo {author} {\bibfnamefont {A.}~\bibnamefont {del
  Campo}},\ }\href {\doibase 10.1103/PhysRevLett.125.040601} {\bibfield
  {journal} {\bibinfo  {journal} {Phys. Rev. Lett.}\ }\textbf {\bibinfo
  {volume} {125}},\ \bibinfo {pages} {040601} (\bibinfo {year}
  {2020})}\BibitemShut {NoStop}%
\bibitem [{\citenamefont {Kamin}\ \emph {et~al.}(2020)\citenamefont {Kamin},
  \citenamefont {Tabesh}, \citenamefont {Salimi},\ and\ \citenamefont
  {Santos}}]{PhysRevE.102.052109}%
  \BibitemOpen
  \bibfield  {author} {\bibinfo {author} {\bibfnamefont {F.~H.}\ \bibnamefont
  {Kamin}}, \bibinfo {author} {\bibfnamefont {F.~T.}\ \bibnamefont {Tabesh}},
  \bibinfo {author} {\bibfnamefont {S.}~\bibnamefont {Salimi}}, \ and\ \bibinfo
  {author} {\bibfnamefont {A.~C.}\ \bibnamefont {Santos}},\ }\href {\doibase
  10.1103/PhysRevE.102.052109} {\bibfield  {journal} {\bibinfo  {journal}
  {Phys. Rev. E}\ }\textbf {\bibinfo {volume} {102}},\ \bibinfo {pages}
  {052109} (\bibinfo {year} {2020})}\BibitemShut {NoStop}%
\bibitem [{\citenamefont {Landi}(2021)}]{e23121627}%
  \BibitemOpen
  \bibfield  {author} {\bibinfo {author} {\bibfnamefont {G.~T.}\ \bibnamefont
  {Landi}},\ }\href {https://www.mdpi.com/1099-4300/23/12/1627} {\bibfield
  {journal} {\bibinfo  {journal} {Entropy}\ }\textbf {\bibinfo {volume} {23}},\
  \bibinfo {pages} {1627} (\bibinfo {year} {2021})}\BibitemShut {NoStop}%
\bibitem [{\citenamefont {Barra}\ \emph {et~al.}(2022)\citenamefont {Barra},
  \citenamefont {Hovhannisyan},\ and\ \citenamefont {Imparato}}]{Barra_2022}%
  \BibitemOpen
  \bibfield  {author} {\bibinfo {author} {\bibfnamefont {F.}~\bibnamefont
  {Barra}}, \bibinfo {author} {\bibfnamefont {K.~V.}\ \bibnamefont
  {Hovhannisyan}}, \ and\ \bibinfo {author} {\bibfnamefont {A.}~\bibnamefont
  {Imparato}},\ }\href {\doibase 10.1088/1367-2630/ac43ed} {\bibfield
  {journal} {\bibinfo  {journal} {New J. Phys.}\ }\textbf {\bibinfo {volume}
  {24}},\ \bibinfo {pages} {015003} (\bibinfo {year} {2022})}\BibitemShut
  {NoStop}%
\bibitem [{\citenamefont {Liu}\ \emph {et~al.}(2021)\citenamefont {Liu},
  \citenamefont {Shi}, \citenamefont {Shi}, \citenamefont {Wang},\ and\
  \citenamefont {Yang}}]{PhysRevB.104.245418}%
  \BibitemOpen
  \bibfield  {author} {\bibinfo {author} {\bibfnamefont {J.-X.}\ \bibnamefont
  {Liu}}, \bibinfo {author} {\bibfnamefont {H.-L.}\ \bibnamefont {Shi}},
  \bibinfo {author} {\bibfnamefont {Y.-H.}\ \bibnamefont {Shi}}, \bibinfo
  {author} {\bibfnamefont {X.-H.}\ \bibnamefont {Wang}}, \ and\ \bibinfo
  {author} {\bibfnamefont {W.-L.}\ \bibnamefont {Yang}},\ }\href {\doibase
  10.1103/PhysRevB.104.245418} {\bibfield  {journal} {\bibinfo  {journal}
  {Phys. Rev. B}\ }\textbf {\bibinfo {volume} {104}},\ \bibinfo {pages}
  {245418} (\bibinfo {year} {2021})}\BibitemShut {NoStop}%
\bibitem [{\citenamefont {Xu}\ \emph {et~al.}(2021)\citenamefont {Xu},
  \citenamefont {Zhu}, \citenamefont {Zhang},\ and\ \citenamefont
  {Liu}}]{PhysRevE.104.064143}%
  \BibitemOpen
  \bibfield  {author} {\bibinfo {author} {\bibfnamefont {K.}~\bibnamefont
  {Xu}}, \bibinfo {author} {\bibfnamefont {H.-J.}\ \bibnamefont {Zhu}},
  \bibinfo {author} {\bibfnamefont {G.-F.}\ \bibnamefont {Zhang}}, \ and\
  \bibinfo {author} {\bibfnamefont {W.-M.}\ \bibnamefont {Liu}},\ }\href
  {\doibase 10.1103/PhysRevE.104.064143} {\bibfield  {journal} {\bibinfo
  {journal} {Phys. Rev. E}\ }\textbf {\bibinfo {volume} {104}},\ \bibinfo
  {pages} {064143} (\bibinfo {year} {2021})}\BibitemShut {NoStop}%
\bibitem [{\citenamefont {Bhattacharjee}\ and\ \citenamefont
  {Dutta}(2021)}]{Bhattacharjee2021}%
  \BibitemOpen
  \bibfield  {author} {\bibinfo {author} {\bibfnamefont {S.}~\bibnamefont
  {Bhattacharjee}}\ and\ \bibinfo {author} {\bibfnamefont {A.}~\bibnamefont
  {Dutta}},\ }\href {\doibase 10.1140/epjb/s10051-021-00235-3} {\bibfield
  {journal} {\bibinfo  {journal} {Eur. Phys. J. B}\ }\textbf {\bibinfo {volume}
  {94}},\ \bibinfo {pages} {239} (\bibinfo {year} {2021})}\BibitemShut
  {NoStop}%
\bibitem [{\citenamefont {Mazzoncini}\ \emph {et~al.}(2023)\citenamefont
  {Mazzoncini}, \citenamefont {Cavina}, \citenamefont {Andolina}, \citenamefont
  {Erdman},\ and\ \citenamefont {Giovannetti}}]{PhysRevA.107.032218}%
  \BibitemOpen
  \bibfield  {author} {\bibinfo {author} {\bibfnamefont {F.}~\bibnamefont
  {Mazzoncini}}, \bibinfo {author} {\bibfnamefont {V.}~\bibnamefont {Cavina}},
  \bibinfo {author} {\bibfnamefont {G.~M.}\ \bibnamefont {Andolina}}, \bibinfo
  {author} {\bibfnamefont {P.~A.}\ \bibnamefont {Erdman}}, \ and\ \bibinfo
  {author} {\bibfnamefont {V.}~\bibnamefont {Giovannetti}},\ }\href {\doibase
  10.1103/PhysRevA.107.032218} {\bibfield  {journal} {\bibinfo  {journal}
  {Phys. Rev. A}\ }\textbf {\bibinfo {volume} {107}},\ \bibinfo {pages}
  {032218} (\bibinfo {year} {2023})}\BibitemShut {NoStop}%
\bibitem [{\citenamefont {Dou}\ \emph {et~al.}(2022{\natexlab{a}})\citenamefont
  {Dou}, \citenamefont {Zhou},\ and\ \citenamefont
  {Sun}}]{PhysRevA.106.032212}%
  \BibitemOpen
  \bibfield  {author} {\bibinfo {author} {\bibfnamefont {F.-Q.}\ \bibnamefont
  {Dou}}, \bibinfo {author} {\bibfnamefont {H.}~\bibnamefont {Zhou}}, \ and\
  \bibinfo {author} {\bibfnamefont {J.-A.}\ \bibnamefont {Sun}},\ }\href
  {\doibase 10.1103/PhysRevA.106.032212} {\bibfield  {journal} {\bibinfo
  {journal} {Phys. Rev. A}\ }\textbf {\bibinfo {volume} {106}},\ \bibinfo
  {pages} {032212} (\bibinfo {year} {2022}{\natexlab{a}})}\BibitemShut
  {NoStop}%
\bibitem [{\citenamefont {Binder}\ \emph {et~al.}(2015)\citenamefont {Binder},
  \citenamefont {Vinjanampathy}, \citenamefont {Modi},\ and\ \citenamefont
  {Goold}}]{Binder_2015}%
  \BibitemOpen
  \bibfield  {author} {\bibinfo {author} {\bibfnamefont {F.~C.}\ \bibnamefont
  {Binder}}, \bibinfo {author} {\bibfnamefont {S.}~\bibnamefont
  {Vinjanampathy}}, \bibinfo {author} {\bibfnamefont {K.}~\bibnamefont {Modi}},
  \ and\ \bibinfo {author} {\bibfnamefont {J.}~\bibnamefont {Goold}},\ }\href
  {\doibase 10.1088/1367-2630/17/7/075015} {\bibfield  {journal} {\bibinfo
  {journal} {New J. Phys}\ }\textbf {\bibinfo {volume} {17}},\ \bibinfo {pages}
  {075015} (\bibinfo {year} {2015})}\BibitemShut {NoStop}%
\bibitem [{\citenamefont {Le}\ \emph {et~al.}(2018)\citenamefont {Le},
  \citenamefont {Levinsen}, \citenamefont {Modi}, \citenamefont {Parish},\ and\
  \citenamefont {Pollock}}]{PhysRevA.97.022106}%
  \BibitemOpen
  \bibfield  {author} {\bibinfo {author} {\bibfnamefont {T.~P.}\ \bibnamefont
  {Le}}, \bibinfo {author} {\bibfnamefont {J.}~\bibnamefont {Levinsen}},
  \bibinfo {author} {\bibfnamefont {K.}~\bibnamefont {Modi}}, \bibinfo {author}
  {\bibfnamefont {M.~M.}\ \bibnamefont {Parish}}, \ and\ \bibinfo {author}
  {\bibfnamefont {F.~A.}\ \bibnamefont {Pollock}},\ }\href {\doibase
  10.1103/PhysRevA.97.022106} {\bibfield  {journal} {\bibinfo  {journal} {Phys.
  Rev. A}\ }\textbf {\bibinfo {volume} {97}},\ \bibinfo {pages} {022106}
  (\bibinfo {year} {2018})}\BibitemShut {NoStop}%
\bibitem [{\citenamefont {Crescente}\ \emph
  {et~al.}(2020{\natexlab{b}})\citenamefont {Crescente}, \citenamefont
  {Carrega}, \citenamefont {Sassetti},\ and\ \citenamefont
  {Ferraro}}]{PhysRevB.102.245407}%
  \BibitemOpen
  \bibfield  {author} {\bibinfo {author} {\bibfnamefont {A.}~\bibnamefont
  {Crescente}}, \bibinfo {author} {\bibfnamefont {M.}~\bibnamefont {Carrega}},
  \bibinfo {author} {\bibfnamefont {M.}~\bibnamefont {Sassetti}}, \ and\
  \bibinfo {author} {\bibfnamefont {D.}~\bibnamefont {Ferraro}},\ }\href
  {\doibase 10.1103/PhysRevB.102.245407} {\bibfield  {journal} {\bibinfo
  {journal} {Phys. Rev. B}\ }\textbf {\bibinfo {volume} {102}},\ \bibinfo
  {pages} {245407} (\bibinfo {year} {2020}{\natexlab{b}})}\BibitemShut
  {NoStop}%
\bibitem [{\citenamefont {Dou}\ \emph {et~al.}(2022{\natexlab{b}})\citenamefont
  {Dou}, \citenamefont {Lu}, \citenamefont {Wang},\ and\ \citenamefont
  {Sun}}]{PhysRevB.105.115405}%
  \BibitemOpen
  \bibfield  {author} {\bibinfo {author} {\bibfnamefont {F.-Q.}\ \bibnamefont
  {Dou}}, \bibinfo {author} {\bibfnamefont {Y.-Q.}\ \bibnamefont {Lu}},
  \bibinfo {author} {\bibfnamefont {Y.-J.}\ \bibnamefont {Wang}}, \ and\
  \bibinfo {author} {\bibfnamefont {J.-A.}\ \bibnamefont {Sun}},\ }\href
  {\doibase 10.1103/PhysRevB.105.115405} {\bibfield  {journal} {\bibinfo
  {journal} {Phys. Rev. B}\ }\textbf {\bibinfo {volume} {105}},\ \bibinfo
  {pages} {115405} (\bibinfo {year} {2022}{\natexlab{b}})}\BibitemShut
  {NoStop}%
\bibitem [{\citenamefont {Yang}\ \emph {et~al.}(2023)\citenamefont {Yang},
  \citenamefont {Yang},\ and\ \citenamefont {Dou}}]{yang2023threelevel}%
  \BibitemOpen
  \bibfield  {author} {\bibinfo {author} {\bibfnamefont {D.-L.}\ \bibnamefont
  {Yang}}, \bibinfo {author} {\bibfnamefont {F.-M.}\ \bibnamefont {Yang}}, \
  and\ \bibinfo {author} {\bibfnamefont {F.-Q.}\ \bibnamefont {Dou}},\
  }\href@noop {} {\enquote {\bibinfo {title} {Three-level dicke quantum
  battery},}\ } (\bibinfo {year} {2023}),\ \Eprint
  {http://arxiv.org/abs/2308.01188} {arXiv:2308.01188 [quant-ph]} \BibitemShut
  {NoStop}%
\bibitem [{\citenamefont {Campaioli}\ \emph {et~al.}(2017)\citenamefont
  {Campaioli}, \citenamefont {Pollock}, \citenamefont {Binder}, \citenamefont
  {C\'eleri}, \citenamefont {Goold}, \citenamefont {Vinjanampathy},\ and\
  \citenamefont {Modi}}]{PhysRevLett.118.150601}%
  \BibitemOpen
  \bibfield  {author} {\bibinfo {author} {\bibfnamefont {F.}~\bibnamefont
  {Campaioli}}, \bibinfo {author} {\bibfnamefont {F.~A.}\ \bibnamefont
  {Pollock}}, \bibinfo {author} {\bibfnamefont {F.~C.}\ \bibnamefont {Binder}},
  \bibinfo {author} {\bibfnamefont {L.}~\bibnamefont {C\'eleri}}, \bibinfo
  {author} {\bibfnamefont {J.}~\bibnamefont {Goold}}, \bibinfo {author}
  {\bibfnamefont {S.}~\bibnamefont {Vinjanampathy}}, \ and\ \bibinfo {author}
  {\bibfnamefont {K.}~\bibnamefont {Modi}},\ }\href {\doibase
  10.1103/PhysRevLett.118.150601} {\bibfield  {journal} {\bibinfo  {journal}
  {Phys. Rev. Lett.}\ }\textbf {\bibinfo {volume} {118}},\ \bibinfo {pages}
  {150601} (\bibinfo {year} {2017})}\BibitemShut {NoStop}%
\bibitem [{\citenamefont {Ferraro}\ \emph {et~al.}(2018)\citenamefont
  {Ferraro}, \citenamefont {Campisi}, \citenamefont {Andolina}, \citenamefont
  {Pellegrini},\ and\ \citenamefont {Polini}}]{PhysRevLett.120.117702}%
  \BibitemOpen
  \bibfield  {author} {\bibinfo {author} {\bibfnamefont {D.}~\bibnamefont
  {Ferraro}}, \bibinfo {author} {\bibfnamefont {M.}~\bibnamefont {Campisi}},
  \bibinfo {author} {\bibfnamefont {G.~M.}\ \bibnamefont {Andolina}}, \bibinfo
  {author} {\bibfnamefont {V.}~\bibnamefont {Pellegrini}}, \ and\ \bibinfo
  {author} {\bibfnamefont {M.}~\bibnamefont {Polini}},\ }\href {\doibase
  10.1103/PhysRevLett.120.117702} {\bibfield  {journal} {\bibinfo  {journal}
  {Phys. Rev. Lett.}\ }\textbf {\bibinfo {volume} {120}},\ \bibinfo {pages}
  {117702} (\bibinfo {year} {2018})}\BibitemShut {NoStop}%
\bibitem [{\citenamefont {Carrega}\ \emph {et~al.}(2020)\citenamefont
  {Carrega}, \citenamefont {Crescente}, \citenamefont {Ferraro},\ and\
  \citenamefont {Sassetti}}]{Carrega_2020}%
  \BibitemOpen
  \bibfield  {author} {\bibinfo {author} {\bibfnamefont {M.}~\bibnamefont
  {Carrega}}, \bibinfo {author} {\bibfnamefont {A.}~\bibnamefont {Crescente}},
  \bibinfo {author} {\bibfnamefont {D.}~\bibnamefont {Ferraro}}, \ and\
  \bibinfo {author} {\bibfnamefont {M.}~\bibnamefont {Sassetti}},\ }\href
  {\doibase 10.1088/1367-2630/abaa01} {\bibfield  {journal} {\bibinfo
  {journal} {New J. Phys.}\ }\textbf {\bibinfo {volume} {22}},\ \bibinfo
  {pages} {083085} (\bibinfo {year} {2020})}\BibitemShut {NoStop}%
\bibitem [{\citenamefont {Rossini}\ \emph {et~al.}(2020)\citenamefont
  {Rossini}, \citenamefont {Andolina}, \citenamefont {Rosa}, \citenamefont
  {Carrega},\ and\ \citenamefont {Polini}}]{PhysRevLett.125.236402}%
  \BibitemOpen
  \bibfield  {author} {\bibinfo {author} {\bibfnamefont {D.}~\bibnamefont
  {Rossini}}, \bibinfo {author} {\bibfnamefont {G.~M.}\ \bibnamefont
  {Andolina}}, \bibinfo {author} {\bibfnamefont {D.}~\bibnamefont {Rosa}},
  \bibinfo {author} {\bibfnamefont {M.}~\bibnamefont {Carrega}}, \ and\
  \bibinfo {author} {\bibfnamefont {M.}~\bibnamefont {Polini}},\ }\href
  {\doibase 10.1103/PhysRevLett.125.236402} {\bibfield  {journal} {\bibinfo
  {journal} {Phys. Rev. Lett.}\ }\textbf {\bibinfo {volume} {125}},\ \bibinfo
  {pages} {236402} (\bibinfo {year} {2020})}\BibitemShut {NoStop}%
\bibitem [{\citenamefont {Llobet}\ and\ \citenamefont
  {Uzdin}(2019)}]{Perarnau_Llobet_2019}%
  \BibitemOpen
  \bibfield  {author} {\bibinfo {author} {\bibfnamefont {M.~P.}\ \bibnamefont
  {Llobet}}\ and\ \bibinfo {author} {\bibfnamefont {R.}~\bibnamefont {Uzdin}},\
  }\href {\doibase 10.1088/1367-2630/ab36a9} {\bibfield  {journal} {\bibinfo
  {journal} {New J. Phys.}\ }\textbf {\bibinfo {volume} {21}},\ \bibinfo
  {pages} {083023} (\bibinfo {year} {2019})}\BibitemShut {NoStop}%
\bibitem [{\citenamefont {Peng}\ \emph {et~al.}(2021)\citenamefont {Peng},
  \citenamefont {He}, \citenamefont {Chesi}, \citenamefont {Lin},\ and\
  \citenamefont {Guan}}]{PhysRevA.103.052220}%
  \BibitemOpen
  \bibfield  {author} {\bibinfo {author} {\bibfnamefont {L.}~\bibnamefont
  {Peng}}, \bibinfo {author} {\bibfnamefont {W.-B.}\ \bibnamefont {He}},
  \bibinfo {author} {\bibfnamefont {S.}~\bibnamefont {Chesi}}, \bibinfo
  {author} {\bibfnamefont {H.-Q.}\ \bibnamefont {Lin}}, \ and\ \bibinfo
  {author} {\bibfnamefont {X.-W.}\ \bibnamefont {Guan}},\ }\href {\doibase
  10.1103/PhysRevA.103.052220} {\bibfield  {journal} {\bibinfo  {journal}
  {Phys. Rev. A}\ }\textbf {\bibinfo {volume} {103}},\ \bibinfo {pages}
  {052220} (\bibinfo {year} {2021})}\BibitemShut {NoStop}%
\bibitem [{\citenamefont {Zakavati}\ \emph {et~al.}(2021)\citenamefont
  {Zakavati}, \citenamefont {Tabesh},\ and\ \citenamefont
  {Salimi}}]{PhysRevE.104.054117}%
  \BibitemOpen
  \bibfield  {author} {\bibinfo {author} {\bibfnamefont {S.}~\bibnamefont
  {Zakavati}}, \bibinfo {author} {\bibfnamefont {F.~T.}\ \bibnamefont
  {Tabesh}}, \ and\ \bibinfo {author} {\bibfnamefont {S.}~\bibnamefont
  {Salimi}},\ }\href {\doibase 10.1103/PhysRevE.104.054117} {\bibfield
  {journal} {\bibinfo  {journal} {Phys. Rev. E}\ }\textbf {\bibinfo {volume}
  {104}},\ \bibinfo {pages} {054117} (\bibinfo {year} {2021})}\BibitemShut
  {NoStop}%
\bibitem [{\citenamefont {Konar}\ \emph {et~al.}(2022)\citenamefont {Konar},
  \citenamefont {Lakkaraju}, \citenamefont {Ghosh},\ and\ \citenamefont
  {Sen(De)}}]{PhysRevA.106.022618}%
  \BibitemOpen
  \bibfield  {author} {\bibinfo {author} {\bibfnamefont {T.~K.}\ \bibnamefont
  {Konar}}, \bibinfo {author} {\bibfnamefont {L.~G.~C.}\ \bibnamefont
  {Lakkaraju}}, \bibinfo {author} {\bibfnamefont {S.}~\bibnamefont {Ghosh}}, \
  and\ \bibinfo {author} {\bibfnamefont {A.}~\bibnamefont {Sen(De)}},\ }\href
  {\doibase 10.1103/PhysRevA.106.022618} {\bibfield  {journal} {\bibinfo
  {journal} {Phys. Rev. A}\ }\textbf {\bibinfo {volume} {106}},\ \bibinfo
  {pages} {022618} (\bibinfo {year} {2022})}\BibitemShut {NoStop}%
\bibitem [{\citenamefont {Friis}\ and\ \citenamefont
  {Huber}(2018)}]{Friis2018precisionwork}%
  \BibitemOpen
  \bibfield  {author} {\bibinfo {author} {\bibfnamefont {N.}~\bibnamefont
  {Friis}}\ and\ \bibinfo {author} {\bibfnamefont {M.}~\bibnamefont {Huber}},\
  }\href {\doibase 10.22331/q-2018-04-23-61} {\bibfield  {journal} {\bibinfo
  {journal} {{Quantum}}\ }\textbf {\bibinfo {volume} {2}},\ \bibinfo {pages}
  {61} (\bibinfo {year} {2018})}\BibitemShut {NoStop}%
\bibitem [{\citenamefont {McKay}\ \emph {et~al.}(2018)\citenamefont {McKay},
  \citenamefont {Rodr\'{\i}guez-Briones},\ and\ \citenamefont
  {Mart\'{\i}n-Mart\'{\i}nez}}]{PhysRevE.98.032132}%
  \BibitemOpen
  \bibfield  {author} {\bibinfo {author} {\bibfnamefont {E.}~\bibnamefont
  {McKay}}, \bibinfo {author} {\bibfnamefont {N.~A.}\ \bibnamefont
  {Rodr\'{\i}guez-Briones}}, \ and\ \bibinfo {author} {\bibfnamefont
  {E.}~\bibnamefont {Mart\'{\i}n-Mart\'{\i}nez}},\ }\href {\doibase
  10.1103/PhysRevE.98.032132} {\bibfield  {journal} {\bibinfo  {journal} {Phys.
  Rev. E}\ }\textbf {\bibinfo {volume} {98}},\ \bibinfo {pages} {032132}
  (\bibinfo {year} {2018})}\BibitemShut {NoStop}%
\bibitem [{\citenamefont {Imai}\ \emph {et~al.}(2023)\citenamefont {Imai},
  \citenamefont {G\"uhne},\ and\ \citenamefont
  {Nimmrichter}}]{PhysRevA.107.022215}%
  \BibitemOpen
  \bibfield  {author} {\bibinfo {author} {\bibfnamefont {S.}~\bibnamefont
  {Imai}}, \bibinfo {author} {\bibfnamefont {O.}~\bibnamefont {G\"uhne}}, \
  and\ \bibinfo {author} {\bibfnamefont {S.}~\bibnamefont {Nimmrichter}},\
  }\href {\doibase 10.1103/PhysRevA.107.022215} {\bibfield  {journal} {\bibinfo
   {journal} {Phys. Rev. A}\ }\textbf {\bibinfo {volume} {107}},\ \bibinfo
  {pages} {022215} (\bibinfo {year} {2023})}\BibitemShut {NoStop}%
\bibitem [{\citenamefont {\ifmmode~\mbox{\c{C}}\else
  \c{C}\fi{}akmak}(2020)}]{PhysRevE.102.042111}%
  \BibitemOpen
  \bibfield  {author} {\bibinfo {author} {\bibfnamefont {B.}~\bibnamefont
  {\ifmmode~\mbox{\c{C}}\else \c{C}\fi{}akmak}},\ }\href {\doibase
  10.1103/PhysRevE.102.042111} {\bibfield  {journal} {\bibinfo  {journal}
  {Phys. Rev. E}\ }\textbf {\bibinfo {volume} {102}},\ \bibinfo {pages}
  {042111} (\bibinfo {year} {2020})}\BibitemShut {NoStop}%
\bibitem [{\citenamefont {Tacchino}\ \emph {et~al.}(2020)\citenamefont
  {Tacchino}, \citenamefont {Santos}, \citenamefont {Gerace}, \citenamefont
  {Campisi},\ and\ \citenamefont {Santos}}]{PhysRevE.102.062133}%
  \BibitemOpen
  \bibfield  {author} {\bibinfo {author} {\bibfnamefont {F.}~\bibnamefont
  {Tacchino}}, \bibinfo {author} {\bibfnamefont {T.~F.~F.}\ \bibnamefont
  {Santos}}, \bibinfo {author} {\bibfnamefont {D.}~\bibnamefont {Gerace}},
  \bibinfo {author} {\bibfnamefont {M.}~\bibnamefont {Campisi}}, \ and\
  \bibinfo {author} {\bibfnamefont {M.~F.}\ \bibnamefont {Santos}},\ }\href
  {\doibase 10.1103/PhysRevE.102.062133} {\bibfield  {journal} {\bibinfo
  {journal} {Phys. Rev. E}\ }\textbf {\bibinfo {volume} {102}},\ \bibinfo
  {pages} {062133} (\bibinfo {year} {2020})}\BibitemShut {NoStop}%
\bibitem [{\citenamefont {Crescente}\ \emph {et~al.}(2023)\citenamefont
  {Crescente}, \citenamefont {Ferraro}, \citenamefont {Carrega},\ and\
  \citenamefont {Sassetti}}]{Crescente_2023}%
  \BibitemOpen
  \bibfield  {author} {\bibinfo {author} {\bibfnamefont {A.}~\bibnamefont
  {Crescente}}, \bibinfo {author} {\bibfnamefont {D.}~\bibnamefont {Ferraro}},
  \bibinfo {author} {\bibfnamefont {M.}~\bibnamefont {Carrega}}, \ and\
  \bibinfo {author} {\bibfnamefont {M.}~\bibnamefont {Sassetti}},\ }\href
  {\doibase 10.3390/e25050758} {\bibfield  {journal} {\bibinfo  {journal}
  {Entropy}\ }\textbf {\bibinfo {volume} {25}},\ \bibinfo {pages} {758}
  (\bibinfo {year} {2023})}\BibitemShut {NoStop}%
\bibitem [{\citenamefont {Chen}\ \emph {et~al.}(2020)\citenamefont {Chen},
  \citenamefont {Zhan}, \citenamefont {Shao}, \citenamefont {Zhang},
  \citenamefont {Zhang},\ and\ \citenamefont {Wang}}]{chen2020}%
  \BibitemOpen
  \bibfield  {author} {\bibinfo {author} {\bibfnamefont {J.}~\bibnamefont
  {Chen}}, \bibinfo {author} {\bibfnamefont {L.}~\bibnamefont {Zhan}}, \bibinfo
  {author} {\bibfnamefont {L.}~\bibnamefont {Shao}}, \bibinfo {author}
  {\bibfnamefont {X.}~\bibnamefont {Zhang}}, \bibinfo {author} {\bibfnamefont
  {Y.}~\bibnamefont {Zhang}}, \ and\ \bibinfo {author} {\bibfnamefont
  {X.}~\bibnamefont {Wang}},\ }\href {\doibase 10.1002/andp.201900487}
  {\bibfield  {journal} {\bibinfo  {journal} {Ann. Phys.}\ }\textbf {\bibinfo
  {volume} {532}},\ \bibinfo {pages} {1900487} (\bibinfo {year}
  {2020})}\BibitemShut {NoStop}%
\bibitem [{\citenamefont {Zhang}\ and\ \citenamefont
  {Blaauboer}(2023)}]{Zhang2023}%
  \BibitemOpen
  \bibfield  {author} {\bibinfo {author} {\bibfnamefont {X.}~\bibnamefont
  {Zhang}}\ and\ \bibinfo {author} {\bibfnamefont {M.}~\bibnamefont
  {Blaauboer}},\ }\href
  {https://www.frontiersin.org/articles/10.3389/fphy.2022.1097564} {\bibfield
  {journal} {\bibinfo  {journal} {Front. Phys.}\ }\textbf {\bibinfo {volume}
  {10}},\ \bibinfo {pages} {1367} (\bibinfo {year} {2023})}\BibitemShut
  {NoStop}%
\bibitem [{\citenamefont {Caravelli}\ \emph {et~al.}(2021)\citenamefont
  {Caravelli}, \citenamefont {Yan}, \citenamefont {Garc{\'{i}}a-Pintos},\ and\
  \citenamefont {Hamma}}]{Caravelli2021energystorage}%
  \BibitemOpen
  \bibfield  {author} {\bibinfo {author} {\bibfnamefont {F.}~\bibnamefont
  {Caravelli}}, \bibinfo {author} {\bibfnamefont {B.}~\bibnamefont {Yan}},
  \bibinfo {author} {\bibfnamefont {L.~P.}\ \bibnamefont
  {Garc{\'{i}}a-Pintos}}, \ and\ \bibinfo {author} {\bibfnamefont
  {A.}~\bibnamefont {Hamma}},\ }\href {\doibase 10.22331/q-2021-07-15-505}
  {\bibfield  {journal} {\bibinfo  {journal} {{Quantum}}\ }\textbf {\bibinfo
  {volume} {5}},\ \bibinfo {pages} {505} (\bibinfo {year} {2021})}\BibitemShut
  {NoStop}%
\bibitem [{\citenamefont {Santos}\ \emph {et~al.}(2019)\citenamefont {Santos},
  \citenamefont {\ifmmode~\mbox{\c{C}}\else \c{C}\fi{}akmak}, \citenamefont
  {Campbell},\ and\ \citenamefont {Zinner}}]{PhysRevE.100.032107}%
  \BibitemOpen
  \bibfield  {author} {\bibinfo {author} {\bibfnamefont {A.~C.}\ \bibnamefont
  {Santos}}, \bibinfo {author} {\bibfnamefont {B.}~\bibnamefont
  {\ifmmode~\mbox{\c{C}}\else \c{C}\fi{}akmak}}, \bibinfo {author}
  {\bibfnamefont {S.}~\bibnamefont {Campbell}}, \ and\ \bibinfo {author}
  {\bibfnamefont {N.~T.}\ \bibnamefont {Zinner}},\ }\href {\doibase
  10.1103/PhysRevE.100.032107} {\bibfield  {journal} {\bibinfo  {journal}
  {Phys. Rev. E}\ }\textbf {\bibinfo {volume} {100}},\ \bibinfo {pages}
  {032107} (\bibinfo {year} {2019})}\BibitemShut {NoStop}%
\bibitem [{\citenamefont {Santos}\ \emph {et~al.}(2020)\citenamefont {Santos},
  \citenamefont {Saguia},\ and\ \citenamefont {Sarandy}}]{PhysRevE.101.062114}%
  \BibitemOpen
  \bibfield  {author} {\bibinfo {author} {\bibfnamefont {A.~C.}\ \bibnamefont
  {Santos}}, \bibinfo {author} {\bibfnamefont {A.}~\bibnamefont {Saguia}}, \
  and\ \bibinfo {author} {\bibfnamefont {M.~S.}\ \bibnamefont {Sarandy}},\
  }\href {\doibase 10.1103/PhysRevE.101.062114} {\bibfield  {journal} {\bibinfo
   {journal} {Phys. Rev. E}\ }\textbf {\bibinfo {volume} {101}},\ \bibinfo
  {pages} {062114} (\bibinfo {year} {2020})}\BibitemShut {NoStop}%
\bibitem [{\citenamefont {Dou}\ \emph {et~al.}(2021)\citenamefont {Dou},
  \citenamefont {Wang},\ and\ \citenamefont {Sun}}]{Dou2021}%
  \BibitemOpen
  \bibfield  {author} {\bibinfo {author} {\bibfnamefont {F.-Q.}\ \bibnamefont
  {Dou}}, \bibinfo {author} {\bibfnamefont {Y.-J.}\ \bibnamefont {Wang}}, \
  and\ \bibinfo {author} {\bibfnamefont {J.-A.}\ \bibnamefont {Sun}},\ }\href
  {\doibase 10.1007/s11467-021-1130-5} {\bibfield  {journal} {\bibinfo
  {journal} {Front. Phys.}\ }\textbf {\bibinfo {volume} {17}},\ \bibinfo
  {pages} {31503} (\bibinfo {year} {2021})}\BibitemShut {NoStop}%
\bibitem [{\citenamefont {Dou}\ \emph {et~al.}(2016{\natexlab{a}})\citenamefont
  {Dou}, \citenamefont {Liu},\ and\ \citenamefont {Fu}}]{Dou_2016}%
  \BibitemOpen
  \bibfield  {author} {\bibinfo {author} {\bibfnamefont {F.-Q.}\ \bibnamefont
  {Dou}}, \bibinfo {author} {\bibfnamefont {J.}~\bibnamefont {Liu}}, \ and\
  \bibinfo {author} {\bibfnamefont {L.-B.}\ \bibnamefont {Fu}},\ }\href
  {\doibase 10.1209/0295-5075/116/60014} {\bibfield  {journal} {\bibinfo
  {journal} {Europhys. Lett.}\ }\textbf {\bibinfo {volume} {116}},\ \bibinfo
  {pages} {60014} (\bibinfo {year} {2016}{\natexlab{a}})}\BibitemShut {NoStop}%
\bibitem [{\citenamefont {Dou}\ \emph {et~al.}(2014)\citenamefont {Dou},
  \citenamefont {Fu},\ and\ \citenamefont {Liu}}]{PhysRevA.89.012123}%
  \BibitemOpen
  \bibfield  {author} {\bibinfo {author} {\bibfnamefont {F.-Q.}\ \bibnamefont
  {Dou}}, \bibinfo {author} {\bibfnamefont {L.-B.}\ \bibnamefont {Fu}}, \ and\
  \bibinfo {author} {\bibfnamefont {J.}~\bibnamefont {Liu}},\ }\href {\doibase
  10.1103/PhysRevA.89.012123} {\bibfield  {journal} {\bibinfo  {journal} {Phys.
  Rev. A}\ }\textbf {\bibinfo {volume} {89}},\ \bibinfo {pages} {012123}
  (\bibinfo {year} {2014})}\BibitemShut {NoStop}%
\bibitem [{\citenamefont {Dou}\ \emph {et~al.}(2016{\natexlab{b}})\citenamefont
  {Dou}, \citenamefont {Cao}, \citenamefont {Liu},\ and\ \citenamefont
  {Fu}}]{PhysRevA.93.043419}%
  \BibitemOpen
  \bibfield  {author} {\bibinfo {author} {\bibfnamefont {F.-Q.}\ \bibnamefont
  {Dou}}, \bibinfo {author} {\bibfnamefont {H.}~\bibnamefont {Cao}}, \bibinfo
  {author} {\bibfnamefont {J.}~\bibnamefont {Liu}}, \ and\ \bibinfo {author}
  {\bibfnamefont {L.-B.}\ \bibnamefont {Fu}},\ }\href {\doibase
  10.1103/PhysRevA.93.043419} {\bibfield  {journal} {\bibinfo  {journal} {Phys.
  Rev. A}\ }\textbf {\bibinfo {volume} {93}},\ \bibinfo {pages} {043419}
  (\bibinfo {year} {2016}{\natexlab{b}})}\BibitemShut {NoStop}%
\bibitem [{\citenamefont {Dou}\ \emph {et~al.}(2018)\citenamefont {Dou},
  \citenamefont {Liu},\ and\ \citenamefont {Fu}}]{PhysRevA.98.022102}%
  \BibitemOpen
  \bibfield  {author} {\bibinfo {author} {\bibfnamefont {F.-Q.}\ \bibnamefont
  {Dou}}, \bibinfo {author} {\bibfnamefont {J.}~\bibnamefont {Liu}}, \ and\
  \bibinfo {author} {\bibfnamefont {L.-B.}\ \bibnamefont {Fu}},\ }\href
  {\doibase 10.1103/PhysRevA.98.022102} {\bibfield  {journal} {\bibinfo
  {journal} {Phys. Rev. A}\ }\textbf {\bibinfo {volume} {98}},\ \bibinfo
  {pages} {022102} (\bibinfo {year} {2018})}\BibitemShut {NoStop}%
\bibitem [{\citenamefont {Hovhannisyan}\ \emph {et~al.}(2013)\citenamefont
  {Hovhannisyan}, \citenamefont {Perarnau-Llobet}, \citenamefont {Huber},\ and\
  \citenamefont {Ac\'{\i}n}}]{PhysRevLett.111.240401}%
  \BibitemOpen
  \bibfield  {author} {\bibinfo {author} {\bibfnamefont {K.~V.}\ \bibnamefont
  {Hovhannisyan}}, \bibinfo {author} {\bibfnamefont {M.}~\bibnamefont
  {Perarnau-Llobet}}, \bibinfo {author} {\bibfnamefont {M.}~\bibnamefont
  {Huber}}, \ and\ \bibinfo {author} {\bibfnamefont {A.}~\bibnamefont
  {Ac\'{\i}n}},\ }\href {\doibase 10.1103/PhysRevLett.111.240401} {\bibfield
  {journal} {\bibinfo  {journal} {Phys. Rev. Lett.}\ }\textbf {\bibinfo
  {volume} {111}},\ \bibinfo {pages} {240401} (\bibinfo {year}
  {2013})}\BibitemShut {NoStop}%
\bibitem [{\citenamefont {Andolina}\ \emph
  {et~al.}(2019{\natexlab{b}})\citenamefont {Andolina}, \citenamefont {Keck},
  \citenamefont {Mari}, \citenamefont {Giovannetti},\ and\ \citenamefont
  {Polini}}]{PhysRevB.99.205437}%
  \BibitemOpen
  \bibfield  {author} {\bibinfo {author} {\bibfnamefont {G.~M.}\ \bibnamefont
  {Andolina}}, \bibinfo {author} {\bibfnamefont {M.}~\bibnamefont {Keck}},
  \bibinfo {author} {\bibfnamefont {A.}~\bibnamefont {Mari}}, \bibinfo {author}
  {\bibfnamefont {V.}~\bibnamefont {Giovannetti}}, \ and\ \bibinfo {author}
  {\bibfnamefont {M.}~\bibnamefont {Polini}},\ }\href {\doibase
  10.1103/PhysRevB.99.205437} {\bibfield  {journal} {\bibinfo  {journal} {Phys.
  Rev. B}\ }\textbf {\bibinfo {volume} {99}},\ \bibinfo {pages} {205437}
  (\bibinfo {year} {2019}{\natexlab{b}})}\BibitemShut {NoStop}%
\bibitem [{\citenamefont {Moraes}\ \emph {et~al.}(2022)\citenamefont {Moraes},
  \citenamefont {Saguia}, \citenamefont {Santos},\ and\ \citenamefont
  {Sarandy}}]{moraes2022charging}%
  \BibitemOpen
  \bibfield  {author} {\bibinfo {author} {\bibfnamefont {L.~F.}\ \bibnamefont
  {Moraes}}, \bibinfo {author} {\bibfnamefont {A.}~\bibnamefont {Saguia}},
  \bibinfo {author} {\bibfnamefont {A.~C.}\ \bibnamefont {Santos}}, \ and\
  \bibinfo {author} {\bibfnamefont {M.~S.}\ \bibnamefont {Sarandy}},\ }\href
  {\doibase 10.1209/0295-5075/ac1363} {\bibfield  {journal} {\bibinfo
  {journal} {Europhys. Lett}\ }\textbf {\bibinfo {volume} {136}},\ \bibinfo
  {pages} {23001} (\bibinfo {year} {2022})}\BibitemShut {NoStop}%
\bibitem [{\citenamefont {Seah}\ \emph {et~al.}(2021)\citenamefont {Seah},
  \citenamefont {Perarnau~Llobet}, \citenamefont {Haack}, \citenamefont
  {Brunner},\ and\ \citenamefont {Nimmrichter}}]{PhysRevLett.127.100601}%
  \BibitemOpen
  \bibfield  {author} {\bibinfo {author} {\bibfnamefont {S.}~\bibnamefont
  {Seah}}, \bibinfo {author} {\bibfnamefont {M.}~\bibnamefont
  {Perarnau~Llobet}}, \bibinfo {author} {\bibfnamefont {G.}~\bibnamefont
  {Haack}}, \bibinfo {author} {\bibfnamefont {N.}~\bibnamefont {Brunner}}, \
  and\ \bibinfo {author} {\bibfnamefont {S.}~\bibnamefont {Nimmrichter}},\
  }\href {\doibase 10.1103/PhysRevLett.127.100601} {\bibfield  {journal}
  {\bibinfo  {journal} {Phys. Rev. Lett.}\ }\textbf {\bibinfo {volume} {127}},\
  \bibinfo {pages} {100601} (\bibinfo {year} {2021})}\BibitemShut {NoStop}%
\bibitem [{\citenamefont {Rossini}\ \emph {et~al.}(2019)\citenamefont
  {Rossini}, \citenamefont {Andolina},\ and\ \citenamefont
  {Polini}}]{PhysRevB.100.115142}%
  \BibitemOpen
  \bibfield  {author} {\bibinfo {author} {\bibfnamefont {D.}~\bibnamefont
  {Rossini}}, \bibinfo {author} {\bibfnamefont {G.~M.}\ \bibnamefont
  {Andolina}}, \ and\ \bibinfo {author} {\bibfnamefont {M.}~\bibnamefont
  {Polini}},\ }\href {\doibase 10.1103/PhysRevB.100.115142} {\bibfield
  {journal} {\bibinfo  {journal} {Phys. Rev. B}\ }\textbf {\bibinfo {volume}
  {100}},\ \bibinfo {pages} {115142} (\bibinfo {year} {2019})}\BibitemShut
  {NoStop}%
\bibitem [{\citenamefont {Fusco}\ \emph {et~al.}(2016)\citenamefont {Fusco},
  \citenamefont {Paternostro},\ and\ \citenamefont
  {De~Chiara}}]{PhysRevE.94.052122}%
  \BibitemOpen
  \bibfield  {author} {\bibinfo {author} {\bibfnamefont {L.}~\bibnamefont
  {Fusco}}, \bibinfo {author} {\bibfnamefont {M.}~\bibnamefont {Paternostro}},
  \ and\ \bibinfo {author} {\bibfnamefont {G.}~\bibnamefont {De~Chiara}},\
  }\href {\doibase 10.1103/PhysRevE.94.052122} {\bibfield  {journal} {\bibinfo
  {journal} {Phys. Rev. E}\ }\textbf {\bibinfo {volume} {94}},\ \bibinfo
  {pages} {052122} (\bibinfo {year} {2016})}\BibitemShut {NoStop}%
\bibitem [{\citenamefont {Quach}\ \emph {et~al.}(2022)\citenamefont {Quach},
  \citenamefont {McGhee}, \citenamefont {Ganzer}, \citenamefont {Rouse},
  \citenamefont {Lovett}, \citenamefont {Gauger}, \citenamefont {Keeling},
  \citenamefont {Cerullo}, \citenamefont {Lidzey},\ and\ \citenamefont
  {Virgili}}]{Superabsorption2022}%
  \BibitemOpen
  \bibfield  {author} {\bibinfo {author} {\bibfnamefont {J.~Q.}\ \bibnamefont
  {Quach}}, \bibinfo {author} {\bibfnamefont {K.~E.}\ \bibnamefont {McGhee}},
  \bibinfo {author} {\bibfnamefont {L.}~\bibnamefont {Ganzer}}, \bibinfo
  {author} {\bibfnamefont {D.~M.}\ \bibnamefont {Rouse}}, \bibinfo {author}
  {\bibfnamefont {B.~W.}\ \bibnamefont {Lovett}}, \bibinfo {author}
  {\bibfnamefont {E.~M.}\ \bibnamefont {Gauger}}, \bibinfo {author}
  {\bibfnamefont {J.}~\bibnamefont {Keeling}}, \bibinfo {author} {\bibfnamefont
  {G.}~\bibnamefont {Cerullo}}, \bibinfo {author} {\bibfnamefont {D.~G.}\
  \bibnamefont {Lidzey}}, \ and\ \bibinfo {author} {\bibfnamefont
  {T.}~\bibnamefont {Virgili}},\ }\href {\doibase 10.1126/sciadv.abk3160}
  {\bibfield  {journal} {\bibinfo  {journal} {Sci. Adv.}\ }\textbf {\bibinfo
  {volume} {8}},\ \bibinfo {pages} {eabk3160} (\bibinfo {year}
  {2022})}\BibitemShut {NoStop}%
\bibitem [{\citenamefont {Mohan}\ and\ \citenamefont
  {Pati}(2021)}]{PhysRevA.104.042209}%
  \BibitemOpen
  \bibfield  {author} {\bibinfo {author} {\bibfnamefont {B.}~\bibnamefont
  {Mohan}}\ and\ \bibinfo {author} {\bibfnamefont {A.~K.}\ \bibnamefont
  {Pati}},\ }\href {\doibase 10.1103/PhysRevA.104.042209} {\bibfield  {journal}
  {\bibinfo  {journal} {Phys. Rev. A}\ }\textbf {\bibinfo {volume} {104}},\
  \bibinfo {pages} {042209} (\bibinfo {year} {2021})}\BibitemShut {NoStop}%
\bibitem [{\citenamefont {Bai}\ and\ \citenamefont
  {An}(2020)}]{PhysRevA.102.060201}%
  \BibitemOpen
  \bibfield  {author} {\bibinfo {author} {\bibfnamefont {S.-Y.}\ \bibnamefont
  {Bai}}\ and\ \bibinfo {author} {\bibfnamefont {J.-H.}\ \bibnamefont {An}},\
  }\href {\doibase 10.1103/PhysRevA.102.060201} {\bibfield  {journal} {\bibinfo
   {journal} {Phys. Rev. A}\ }\textbf {\bibinfo {volume} {102}},\ \bibinfo
  {pages} {060201} (\bibinfo {year} {2020})}\BibitemShut {NoStop}%
\bibitem [{\citenamefont {Rosen}\ and\ \citenamefont
  {Zener}(1932)}]{PhysRev.40.502}%
  \BibitemOpen
  \bibfield  {author} {\bibinfo {author} {\bibfnamefont {N.}~\bibnamefont
  {Rosen}}\ and\ \bibinfo {author} {\bibfnamefont {C.}~\bibnamefont {Zener}},\
  }\href {\doibase 10.1103/PhysRev.40.502} {\bibfield  {journal} {\bibinfo
  {journal} {Phys. Rev.}\ }\textbf {\bibinfo {volume} {40}},\ \bibinfo {pages}
  {502} (\bibinfo {year} {1932})}\BibitemShut {NoStop}%
\bibitem [{\citenamefont {Torosov}\ and\ \citenamefont
  {Vitanov}(2007)}]{PhysRevA.76.053404}%
  \BibitemOpen
  \bibfield  {author} {\bibinfo {author} {\bibfnamefont {B.~T.}\ \bibnamefont
  {Torosov}}\ and\ \bibinfo {author} {\bibfnamefont {N.~V.}\ \bibnamefont
  {Vitanov}},\ }\href {\doibase 10.1103/PhysRevA.76.053404} {\bibfield
  {journal} {\bibinfo  {journal} {Phys. Rev. A}\ }\textbf {\bibinfo {volume}
  {76}},\ \bibinfo {pages} {053404} (\bibinfo {year} {2007})}\BibitemShut
  {NoStop}%
\bibitem [{\citenamefont {Xu}\ \emph {et~al.}(2008)\citenamefont {Xu},
  \citenamefont {Lu},\ and\ \citenamefont {Li}}]{PhysRevA.78.043609}%
  \BibitemOpen
  \bibfield  {author} {\bibinfo {author} {\bibfnamefont {X.-Q.}\ \bibnamefont
  {Xu}}, \bibinfo {author} {\bibfnamefont {L.-H.}\ \bibnamefont {Lu}}, \ and\
  \bibinfo {author} {\bibfnamefont {Y.-Q.}\ \bibnamefont {Li}},\ }\href
  {\doibase 10.1103/PhysRevA.78.043609} {\bibfield  {journal} {\bibinfo
  {journal} {Phys. Rev. A}\ }\textbf {\bibinfo {volume} {78}},\ \bibinfo
  {pages} {043609} (\bibinfo {year} {2008})}\BibitemShut {NoStop}%
\bibitem [{\citenamefont {Ishkhanyan}\ \emph {et~al.}(2009)\citenamefont
  {Ishkhanyan}, \citenamefont {Sokhoyan}, \citenamefont {Joulakian},\ and\
  \citenamefont {Suominen}}]{ISHKHANYAN2009218}%
  \BibitemOpen
  \bibfield  {author} {\bibinfo {author} {\bibfnamefont {A.}~\bibnamefont
  {Ishkhanyan}}, \bibinfo {author} {\bibfnamefont {R.}~\bibnamefont
  {Sokhoyan}}, \bibinfo {author} {\bibfnamefont {B.}~\bibnamefont {Joulakian}},
  \ and\ \bibinfo {author} {\bibfnamefont {K.-A.}\ \bibnamefont {Suominen}},\
  }\href {\doibase https://doi.org/10.1016/j.optcom.2008.10.008} {\bibfield
  {journal} {\bibinfo  {journal} {Opt. Commun.}\ }\textbf {\bibinfo {volume}
  {282}},\ \bibinfo {pages} {218} (\bibinfo {year} {2009})}\BibitemShut
  {NoStop}%
\bibitem [{\citenamefont {Fu}\ \emph {et~al.}(2009)\citenamefont {Fu},
  \citenamefont {Ye}, \citenamefont {Lee}, \citenamefont {Zhang},\ and\
  \citenamefont {Liu}}]{PhysRevA.80.013619}%
  \BibitemOpen
  \bibfield  {author} {\bibinfo {author} {\bibfnamefont {L.-B.}\ \bibnamefont
  {Fu}}, \bibinfo {author} {\bibfnamefont {D.-F.}\ \bibnamefont {Ye}}, \bibinfo
  {author} {\bibfnamefont {C.}~\bibnamefont {Lee}}, \bibinfo {author}
  {\bibfnamefont {W.}~\bibnamefont {Zhang}}, \ and\ \bibinfo {author}
  {\bibfnamefont {J.}~\bibnamefont {Liu}},\ }\href {\doibase
  10.1103/PhysRevA.80.013619} {\bibfield  {journal} {\bibinfo  {journal} {Phys.
  Rev. A}\ }\textbf {\bibinfo {volume} {80}},\ \bibinfo {pages} {013619}
  (\bibinfo {year} {2009})}\BibitemShut {NoStop}%
\bibitem [{\citenamefont {Li}\ \emph {et~al.}(2008)\citenamefont {Li},
  \citenamefont {Fu}, \citenamefont {Duan},\ and\ \citenamefont
  {Liu}}]{PhysRevA.78.063621}%
  \BibitemOpen
  \bibfield  {author} {\bibinfo {author} {\bibfnamefont {S.-C.}\ \bibnamefont
  {Li}}, \bibinfo {author} {\bibfnamefont {L.-B.}\ \bibnamefont {Fu}}, \bibinfo
  {author} {\bibfnamefont {W.-S.}\ \bibnamefont {Duan}}, \ and\ \bibinfo
  {author} {\bibfnamefont {J.}~\bibnamefont {Liu}},\ }\href {\doibase
  10.1103/PhysRevA.78.063621} {\bibfield  {journal} {\bibinfo  {journal} {Phys.
  Rev. A}\ }\textbf {\bibinfo {volume} {78}},\ \bibinfo {pages} {063621}
  (\bibinfo {year} {2008})}\BibitemShut {NoStop}%
\bibitem [{\citenamefont {Li}\ and\ \citenamefont
  {Fu}(2020{\natexlab{a}})}]{PhysRevA.101.023618}%
  \BibitemOpen
  \bibfield  {author} {\bibinfo {author} {\bibfnamefont {S.-C.}\ \bibnamefont
  {Li}}\ and\ \bibinfo {author} {\bibfnamefont {L.-B.}\ \bibnamefont {Fu}},\
  }\href {\doibase 10.1103/PhysRevA.101.023618} {\bibfield  {journal} {\bibinfo
   {journal} {Phys. Rev. A}\ }\textbf {\bibinfo {volume} {101}},\ \bibinfo
  {pages} {023618} (\bibinfo {year} {2020}{\natexlab{a}})}\BibitemShut
  {NoStop}%
\bibitem [{\citenamefont {Li}\ and\ \citenamefont
  {Fu}(2020{\natexlab{b}})}]{PhysRevA.102.033323}%
  \BibitemOpen
  \bibfield  {author} {\bibinfo {author} {\bibfnamefont {S.-C.}\ \bibnamefont
  {Li}}\ and\ \bibinfo {author} {\bibfnamefont {L.-B.}\ \bibnamefont {Fu}},\
  }\href {\doibase 10.1103/PhysRevA.102.033323} {\bibfield  {journal} {\bibinfo
   {journal} {Phys. Rev. A}\ }\textbf {\bibinfo {volume} {102}},\ \bibinfo
  {pages} {033323} (\bibinfo {year} {2020}{\natexlab{b}})}\BibitemShut
  {NoStop}%
\bibitem [{\citenamefont {Klich}\ \emph {et~al.}(2007)\citenamefont {Klich},
  \citenamefont {Lannert},\ and\ \citenamefont
  {Refael}}]{PhysRevLett.99.205303}%
  \BibitemOpen
  \bibfield  {author} {\bibinfo {author} {\bibfnamefont {I.}~\bibnamefont
  {Klich}}, \bibinfo {author} {\bibfnamefont {C.}~\bibnamefont {Lannert}}, \
  and\ \bibinfo {author} {\bibfnamefont {G.}~\bibnamefont {Refael}},\ }\href
  {\doibase 10.1103/PhysRevLett.99.205303} {\bibfield  {journal} {\bibinfo
  {journal} {Phys. Rev. Lett.}\ }\textbf {\bibinfo {volume} {99}},\ \bibinfo
  {pages} {205303} (\bibinfo {year} {2007})}\BibitemShut {NoStop}%
\bibitem [{\citenamefont {Campbell}\ \emph {et~al.}(2010)\citenamefont
  {Campbell}, \citenamefont {Mizrahi}, \citenamefont {Quraishi}, \citenamefont
  {Senko}, \citenamefont {Hayes}, \citenamefont {Hucul}, \citenamefont
  {Matsukevich}, \citenamefont {Maunz},\ and\ \citenamefont
  {Monroe}}]{PhysRevLett.105.090502}%
  \BibitemOpen
  \bibfield  {author} {\bibinfo {author} {\bibfnamefont {W.~C.}\ \bibnamefont
  {Campbell}}, \bibinfo {author} {\bibfnamefont {J.}~\bibnamefont {Mizrahi}},
  \bibinfo {author} {\bibfnamefont {Q.}~\bibnamefont {Quraishi}}, \bibinfo
  {author} {\bibfnamefont {C.}~\bibnamefont {Senko}}, \bibinfo {author}
  {\bibfnamefont {D.}~\bibnamefont {Hayes}}, \bibinfo {author} {\bibfnamefont
  {D.}~\bibnamefont {Hucul}}, \bibinfo {author} {\bibfnamefont {D.~N.}\
  \bibnamefont {Matsukevich}}, \bibinfo {author} {\bibfnamefont
  {P.}~\bibnamefont {Maunz}}, \ and\ \bibinfo {author} {\bibfnamefont
  {C.}~\bibnamefont {Monroe}},\ }\href {\doibase
  10.1103/PhysRevLett.105.090502} {\bibfield  {journal} {\bibinfo  {journal}
  {Phys. Rev. Lett.}\ }\textbf {\bibinfo {volume} {105}},\ \bibinfo {pages}
  {090502} (\bibinfo {year} {2010})}\BibitemShut {NoStop}%
\bibitem [{\citenamefont {Ye}\ \emph {et~al.}(2008)\citenamefont {Ye},
  \citenamefont {Fu},\ and\ \citenamefont {Liu}}]{PhysRevA.77.013402}%
  \BibitemOpen
  \bibfield  {author} {\bibinfo {author} {\bibfnamefont {D.-F.}\ \bibnamefont
  {Ye}}, \bibinfo {author} {\bibfnamefont {L.-B.}\ \bibnamefont {Fu}}, \ and\
  \bibinfo {author} {\bibfnamefont {J.}~\bibnamefont {Liu}},\ }\href {\doibase
  10.1103/PhysRevA.77.013402} {\bibfield  {journal} {\bibinfo  {journal} {Phys.
  Rev. A}\ }\textbf {\bibinfo {volume} {77}},\ \bibinfo {pages} {013402}
  (\bibinfo {year} {2008})}\BibitemShut {NoStop}%
\bibitem [{\citenamefont {Gemme}\ \emph {et~al.}(2022)\citenamefont {Gemme},
  \citenamefont {Grossi}, \citenamefont {Ferraro}, \citenamefont {Vallecorsa},\
  and\ \citenamefont {Sassetti}}]{batteries8050043}%
  \BibitemOpen
  \bibfield  {author} {\bibinfo {author} {\bibfnamefont {G.}~\bibnamefont
  {Gemme}}, \bibinfo {author} {\bibfnamefont {M.}~\bibnamefont {Grossi}},
  \bibinfo {author} {\bibfnamefont {D.}~\bibnamefont {Ferraro}}, \bibinfo
  {author} {\bibfnamefont {S.}~\bibnamefont {Vallecorsa}}, \ and\ \bibinfo
  {author} {\bibfnamefont {M.}~\bibnamefont {Sassetti}},\ }\href
  {https://www.mdpi.com/2313-0105/8/5/43} {\bibfield  {journal} {\bibinfo
  {journal} {Batteries}\ }\textbf {\bibinfo {volume} {8}},\ \bibinfo {pages}
  {43} (\bibinfo {year} {2022})}\BibitemShut {NoStop}%
\bibitem [{\citenamefont {Wang}\ \emph {et~al.}(1993)\citenamefont {Wang},
  \citenamefont {Li},\ and\ \citenamefont {Weiguny}}]{WANG1993189}%
  \BibitemOpen
  \bibfield  {author} {\bibinfo {author} {\bibfnamefont {S.}~\bibnamefont
  {Wang}}, \bibinfo {author} {\bibfnamefont {F.}~\bibnamefont {Li}}, \ and\
  \bibinfo {author} {\bibfnamefont {A.}~\bibnamefont {Weiguny}},\ }\href
  {\doibase https://doi.org/10.1016/0375-9601(93)90694-U} {\bibfield  {journal}
  {\bibinfo  {journal} {Physics Letters A}\ }\textbf {\bibinfo {volume}
  {180}},\ \bibinfo {pages} {189} (\bibinfo {year} {1993})}\BibitemShut
  {NoStop}%
\bibitem [{\citenamefont {Wang}\ and\ \citenamefont {Zuo}(1994)}]{WANG199413}%
  \BibitemOpen
  \bibfield  {author} {\bibinfo {author} {\bibfnamefont {S.}~\bibnamefont
  {Wang}}\ and\ \bibinfo {author} {\bibfnamefont {W.}~\bibnamefont {Zuo}},\
  }\href {\doibase https://doi.org/10.1016/0375-9601(94)91035-9} {\bibfield
  {journal} {\bibinfo  {journal} {Physics Letters A}\ }\textbf {\bibinfo
  {volume} {196}},\ \bibinfo {pages} {13} (\bibinfo {year} {1994})}\BibitemShut
  {NoStop}%
\bibitem [{\citenamefont {Zhao}\ \emph {et~al.}(2018)\citenamefont {Zhao},
  \citenamefont {Li}, \citenamefont {Cao}, \citenamefont {Yao},\ and\
  \citenamefont {Cen}}]{PhysRevA.98.022136}%
  \BibitemOpen
  \bibfield  {author} {\bibinfo {author} {\bibfnamefont {P.-J.}\ \bibnamefont
  {Zhao}}, \bibinfo {author} {\bibfnamefont {W.}~\bibnamefont {Li}}, \bibinfo
  {author} {\bibfnamefont {H.}~\bibnamefont {Cao}}, \bibinfo {author}
  {\bibfnamefont {S.-W.}\ \bibnamefont {Yao}}, \ and\ \bibinfo {author}
  {\bibfnamefont {L.-X.}\ \bibnamefont {Cen}},\ }\href {\doibase
  10.1103/PhysRevA.98.022136} {\bibfield  {journal} {\bibinfo  {journal} {Phys.
  Rev. A}\ }\textbf {\bibinfo {volume} {98}},\ \bibinfo {pages} {022136}
  (\bibinfo {year} {2018})}\BibitemShut {NoStop}%
\bibitem [{\citenamefont {Thomas}(1983)}]{PhysRevA.27.2744}%
  \BibitemOpen
  \bibfield  {author} {\bibinfo {author} {\bibfnamefont {G.~F.}\ \bibnamefont
  {Thomas}},\ }\href {\doibase 10.1103/PhysRevA.27.2744} {\bibfield  {journal}
  {\bibinfo  {journal} {Phys. Rev. A}\ }\textbf {\bibinfo {volume} {27}},\
  \bibinfo {pages} {2744} (\bibinfo {year} {1983})}\BibitemShut {NoStop}%
\bibitem [{\citenamefont {Osherov}\ and\ \citenamefont
  {Voronin}(1994)}]{PhysRevA.49.265}%
  \BibitemOpen
  \bibfield  {author} {\bibinfo {author} {\bibfnamefont {V.~I.}\ \bibnamefont
  {Osherov}}\ and\ \bibinfo {author} {\bibfnamefont {A.~I.}\ \bibnamefont
  {Voronin}},\ }\href {\doibase 10.1103/PhysRevA.49.265} {\bibfield  {journal}
  {\bibinfo  {journal} {Phys. Rev. A}\ }\textbf {\bibinfo {volume} {49}},\
  \bibinfo {pages} {265} (\bibinfo {year} {1994})}\BibitemShut {NoStop}%
\bibitem [{\citenamefont {Simeonov}\ and\ \citenamefont
  {Vitanov}(2014)}]{PhysRevA.89.043411}%
  \BibitemOpen
  \bibfield  {author} {\bibinfo {author} {\bibfnamefont {L.~S.}\ \bibnamefont
  {Simeonov}}\ and\ \bibinfo {author} {\bibfnamefont {N.~V.}\ \bibnamefont
  {Vitanov}},\ }\href {\doibase 10.1103/PhysRevA.89.043411} {\bibfield
  {journal} {\bibinfo  {journal} {Phys. Rev. A}\ }\textbf {\bibinfo {volume}
  {89}},\ \bibinfo {pages} {043411} (\bibinfo {year} {2014})}\BibitemShut
  {NoStop}%
\bibitem [{\citenamefont {Dutta}\ \emph {et~al.}(2015)\citenamefont {Dutta},
  \citenamefont {Aeppli}, \citenamefont {Chakrabarti}, \citenamefont
  {Divakaran}, \citenamefont {Rosenbaum},\ and\ \citenamefont
  {Sen}}]{dutta_aeppli_chakrabarti_divakaran_rosenbaum_sen_2015}%
  \BibitemOpen
  \bibfield  {author} {\bibinfo {author} {\bibfnamefont {A.}~\bibnamefont
  {Dutta}}, \bibinfo {author} {\bibfnamefont {G.}~\bibnamefont {Aeppli}},
  \bibinfo {author} {\bibfnamefont {B.~K.}\ \bibnamefont {Chakrabarti}},
  \bibinfo {author} {\bibfnamefont {U.}~\bibnamefont {Divakaran}}, \bibinfo
  {author} {\bibfnamefont {T.~F.}\ \bibnamefont {Rosenbaum}}, \ and\ \bibinfo
  {author} {\bibfnamefont {D.}~\bibnamefont {Sen}},\ }\href {\doibase
  10.1017/CBO9781107706057} {\emph {\bibinfo {title} {Quantum Phase Transitions
  in Transverse Field Spin Models: From Statistical Physics to Quantum
  Information}}}\ (\bibinfo  {publisher} {Cambridge University Press},\
  \bibinfo {year} {2015})\BibitemShut {NoStop}%
\bibitem [{\citenamefont {Baumgratz}\ \emph {et~al.}(2014)\citenamefont
  {Baumgratz}, \citenamefont {Cramer},\ and\ \citenamefont
  {Plenio}}]{PhysRevLett.113.140401}%
  \BibitemOpen
  \bibfield  {author} {\bibinfo {author} {\bibfnamefont {T.}~\bibnamefont
  {Baumgratz}}, \bibinfo {author} {\bibfnamefont {M.}~\bibnamefont {Cramer}}, \
  and\ \bibinfo {author} {\bibfnamefont {M.~B.}\ \bibnamefont {Plenio}},\
  }\href {\doibase 10.1103/PhysRevLett.113.140401} {\bibfield  {journal}
  {\bibinfo  {journal} {Phys. Rev. Lett.}\ }\textbf {\bibinfo {volume} {113}},\
  \bibinfo {pages} {140401} (\bibinfo {year} {2014})}\BibitemShut {NoStop}%
\bibitem [{\citenamefont {Shi}\ \emph {et~al.}(2022)\citenamefont {Shi},
  \citenamefont {Ding}, \citenamefont {Wan}, \citenamefont {Wang},\ and\
  \citenamefont {Yang}}]{PhysRevLett.129.130602}%
  \BibitemOpen
  \bibfield  {author} {\bibinfo {author} {\bibfnamefont {H.-L.}\ \bibnamefont
  {Shi}}, \bibinfo {author} {\bibfnamefont {S.}~\bibnamefont {Ding}}, \bibinfo
  {author} {\bibfnamefont {Q.-K.}\ \bibnamefont {Wan}}, \bibinfo {author}
  {\bibfnamefont {X.-H.}\ \bibnamefont {Wang}}, \ and\ \bibinfo {author}
  {\bibfnamefont {W.-L.}\ \bibnamefont {Yang}},\ }\href {\doibase
  10.1103/PhysRevLett.129.130602} {\bibfield  {journal} {\bibinfo  {journal}
  {Phys. Rev. Lett.}\ }\textbf {\bibinfo {volume} {129}},\ \bibinfo {pages}
  {130602} (\bibinfo {year} {2022})}\BibitemShut {NoStop}%
\bibitem [{\citenamefont {Li}\ and\ \citenamefont
  {Cen}(2018{\natexlab{a}})}]{LI20181}%
  \BibitemOpen
  \bibfield  {author} {\bibinfo {author} {\bibfnamefont {W.}~\bibnamefont
  {Li}}\ and\ \bibinfo {author} {\bibfnamefont {L.-X.}\ \bibnamefont {Cen}},\
  }\href {\doibase https://doi.org/10.1016/j.aop.2017.12.002} {\bibfield
  {journal} {\bibinfo  {journal} {Ann. Phys.}\ }\textbf {\bibinfo {volume}
  {389}},\ \bibinfo {pages} {1} (\bibinfo {year}
  {2018}{\natexlab{a}})}\BibitemShut {NoStop}%
\bibitem [{\citenamefont {Li}\ and\ \citenamefont
  {Cen}(2018{\natexlab{b}})}]{Li2018}%
  \BibitemOpen
  \bibfield  {author} {\bibinfo {author} {\bibfnamefont {W.}~\bibnamefont
  {Li}}\ and\ \bibinfo {author} {\bibfnamefont {L.-X.}\ \bibnamefont {Cen}},\
  }\href {\doibase 10.1007/s11128-018-1869-y} {\bibfield  {journal} {\bibinfo
  {journal} {Quantum Inf. Process.}\ }\textbf {\bibinfo {volume} {17}},\
  \bibinfo {pages} {97} (\bibinfo {year} {2018}{\natexlab{b}})}\BibitemShut
  {NoStop}%
\bibitem [{\citenamefont {Wang}\ and\ \citenamefont
  {Cen}(1998)}]{PhysRevA.58.3328}%
  \BibitemOpen
  \bibfield  {author} {\bibinfo {author} {\bibfnamefont {S.-J.}\ \bibnamefont
  {Wang}}\ and\ \bibinfo {author} {\bibfnamefont {L.-X.}\ \bibnamefont {Cen}},\
  }\href {\doibase 10.1103/PhysRevA.58.3328} {\bibfield  {journal} {\bibinfo
  {journal} {Phys. Rev. A}\ }\textbf {\bibinfo {volume} {58}},\ \bibinfo
  {pages} {3328} (\bibinfo {year} {1998})}\BibitemShut {NoStop}%
\bibitem [{\citenamefont {Evangelakos}\ \emph {et~al.}(2023)\citenamefont
  {Evangelakos}, \citenamefont {Paspalakis},\ and\ \citenamefont
  {Stefanatos}}]{PhysRevA.108.062425}%
  \BibitemOpen
  \bibfield  {author} {\bibinfo {author} {\bibfnamefont {V.}~\bibnamefont
  {Evangelakos}}, \bibinfo {author} {\bibfnamefont {E.}~\bibnamefont
  {Paspalakis}}, \ and\ \bibinfo {author} {\bibfnamefont {D.}~\bibnamefont
  {Stefanatos}},\ }\href {\doibase 10.1103/PhysRevA.108.062425} {\bibfield
  {journal} {\bibinfo  {journal} {Phys. Rev. A}\ }\textbf {\bibinfo {volume}
  {108}},\ \bibinfo {pages} {062425} (\bibinfo {year} {2023})}\BibitemShut
  {NoStop}%
\bibitem [{\citenamefont {Joshi}\ and\ \citenamefont
  {Mahesh}(2022)}]{PhysRevA.106.042601}%
  \BibitemOpen
  \bibfield  {author} {\bibinfo {author} {\bibfnamefont {J.}~\bibnamefont
  {Joshi}}\ and\ \bibinfo {author} {\bibfnamefont {T.~S.}\ \bibnamefont
  {Mahesh}},\ }\href {\doibase 10.1103/PhysRevA.106.042601} {\bibfield
  {journal} {\bibinfo  {journal} {Phys. Rev. A}\ }\textbf {\bibinfo {volume}
  {106}},\ \bibinfo {pages} {042601} (\bibinfo {year} {2022})}\BibitemShut
  {NoStop}%
\bibitem [{\citenamefont {Wenniger}\ \emph {et~al.}()\citenamefont {Wenniger},
  \citenamefont {Thomas}, \citenamefont {Maffei}, \citenamefont {Wein},
  \citenamefont {Pont}, \citenamefont {Harouri}, \citenamefont {Lema{\^\i}tre},
  \citenamefont {Sagnes}, \citenamefont {Somaschi}, \citenamefont
  {Auff{\`e}ves} \emph {et~al.}}]{wenniger2022coherence}%
  \BibitemOpen
  \bibfield  {author} {\bibinfo {author} {\bibfnamefont {I.}~\bibnamefont
  {Wenniger}}, \bibinfo {author} {\bibfnamefont {S.}~\bibnamefont {Thomas}},
  \bibinfo {author} {\bibfnamefont {M.}~\bibnamefont {Maffei}}, \bibinfo
  {author} {\bibfnamefont {S.}~\bibnamefont {Wein}}, \bibinfo {author}
  {\bibfnamefont {M.}~\bibnamefont {Pont}}, \bibinfo {author} {\bibfnamefont
  {A.}~\bibnamefont {Harouri}}, \bibinfo {author} {\bibfnamefont
  {A.}~\bibnamefont {Lema{\^\i}tre}}, \bibinfo {author} {\bibfnamefont
  {I.}~\bibnamefont {Sagnes}}, \bibinfo {author} {\bibfnamefont
  {N.}~\bibnamefont {Somaschi}}, \bibinfo {author} {\bibfnamefont
  {A.}~\bibnamefont {Auff{\`e}ves}},  \emph {et~al.},\ }\href@noop {} {}\Eprint
  {http://arxiv.org/abs/arXiv: 2202. 01109} {arXiv: 2202. 01109} \BibitemShut
  {NoStop}%
\bibitem [{\citenamefont {Hu}\ \emph {et~al.}(2022)\citenamefont {Hu},
  \citenamefont {Qiu}, \citenamefont {Souza}, \citenamefont {Yuan},
  \citenamefont {Zhou}, \citenamefont {Zhang}, \citenamefont {Chu},
  \citenamefont {Pan}, \citenamefont {Hu}, \citenamefont {Li}, \citenamefont
  {Xu}, \citenamefont {Zhong}, \citenamefont {Liu}, \citenamefont {Yan},
  \citenamefont {Tan}, \citenamefont {Bachelard}, \citenamefont {Villas-Boas},
  \citenamefont {Santos},\ and\ \citenamefont {Yu}}]{Hu_2022}%
  \BibitemOpen
  \bibfield  {author} {\bibinfo {author} {\bibfnamefont {C.-K.}\ \bibnamefont
  {Hu}}, \bibinfo {author} {\bibfnamefont {J.}~\bibnamefont {Qiu}}, \bibinfo
  {author} {\bibfnamefont {P.~J.~P.}\ \bibnamefont {Souza}}, \bibinfo {author}
  {\bibfnamefont {J.}~\bibnamefont {Yuan}}, \bibinfo {author} {\bibfnamefont
  {Y.}~\bibnamefont {Zhou}}, \bibinfo {author} {\bibfnamefont {L.}~\bibnamefont
  {Zhang}}, \bibinfo {author} {\bibfnamefont {J.}~\bibnamefont {Chu}}, \bibinfo
  {author} {\bibfnamefont {X.}~\bibnamefont {Pan}}, \bibinfo {author}
  {\bibfnamefont {L.}~\bibnamefont {Hu}}, \bibinfo {author} {\bibfnamefont
  {J.}~\bibnamefont {Li}}, \bibinfo {author} {\bibfnamefont {Y.}~\bibnamefont
  {Xu}}, \bibinfo {author} {\bibfnamefont {Y.}~\bibnamefont {Zhong}}, \bibinfo
  {author} {\bibfnamefont {S.}~\bibnamefont {Liu}}, \bibinfo {author}
  {\bibfnamefont {F.}~\bibnamefont {Yan}}, \bibinfo {author} {\bibfnamefont
  {D.}~\bibnamefont {Tan}}, \bibinfo {author} {\bibfnamefont {R.}~\bibnamefont
  {Bachelard}}, \bibinfo {author} {\bibfnamefont {C.~J.}\ \bibnamefont
  {Villas-Boas}}, \bibinfo {author} {\bibfnamefont {A.~C.}\ \bibnamefont
  {Santos}}, \ and\ \bibinfo {author} {\bibfnamefont {D.}~\bibnamefont {Yu}},\
  }\href {\doibase 10.1088/2058-9565/ac8444} {\bibfield  {journal} {\bibinfo
  {journal} {Quantum Sci. Technol.}\ }\textbf {\bibinfo {volume} {7}},\
  \bibinfo {pages} {045018} (\bibinfo {year} {2022})}\BibitemShut {NoStop}%
\bibitem [{\citenamefont {Dou}\ and\ \citenamefont
  {Yang}(2023)}]{PhysRevA.107.023725}%
  \BibitemOpen
  \bibfield  {author} {\bibinfo {author} {\bibfnamefont {F.-Q.}\ \bibnamefont
  {Dou}}\ and\ \bibinfo {author} {\bibfnamefont {F.-M.}\ \bibnamefont {Yang}},\
  }\href {\doibase 10.1103/PhysRevA.107.023725} {\bibfield  {journal} {\bibinfo
   {journal} {Phys. Rev. A}\ }\textbf {\bibinfo {volume} {107}},\ \bibinfo
  {pages} {023725} (\bibinfo {year} {2023})}\BibitemShut {NoStop}%
\bibitem [{\citenamefont {Zheng}\ \emph {et~al.}(2022)\citenamefont {Zheng},
  \citenamefont {Ning}, \citenamefont {Yang}, \citenamefont {Xia},\ and\
  \citenamefont {Zheng}}]{Zheng_2022}%
  \BibitemOpen
  \bibfield  {author} {\bibinfo {author} {\bibfnamefont {R.-H.}\ \bibnamefont
  {Zheng}}, \bibinfo {author} {\bibfnamefont {W.}~\bibnamefont {Ning}},
  \bibinfo {author} {\bibfnamefont {Z.-B.}\ \bibnamefont {Yang}}, \bibinfo
  {author} {\bibfnamefont {Y.}~\bibnamefont {Xia}}, \ and\ \bibinfo {author}
  {\bibfnamefont {S.-B.}\ \bibnamefont {Zheng}},\ }\href {\doibase
  10.1088/1367-2630/ac788f} {\bibfield  {journal} {\bibinfo  {journal} {New J.
  Phys.}\ }\textbf {\bibinfo {volume} {24}},\ \bibinfo {pages} {063031}
  (\bibinfo {year} {2022})}\BibitemShut {NoStop}%
\bibitem [{\citenamefont {Huang}\ \emph {et~al.}(2023)\citenamefont {Huang},
  \citenamefont {Wang}, \citenamefont {Xiao}, \citenamefont {Gao},
  \citenamefont {Lin},\ and\ \citenamefont {Xue}}]{PhysRevA.107.L030201}%
  \BibitemOpen
  \bibfield  {author} {\bibinfo {author} {\bibfnamefont {X.-J.}\ \bibnamefont
  {Huang}}, \bibinfo {author} {\bibfnamefont {K.-K.}\ \bibnamefont {Wang}},
  \bibinfo {author} {\bibfnamefont {L.}~\bibnamefont {Xiao}}, \bibinfo {author}
  {\bibfnamefont {L.}~\bibnamefont {Gao}}, \bibinfo {author} {\bibfnamefont
  {H.-Q.}\ \bibnamefont {Lin}}, \ and\ \bibinfo {author} {\bibfnamefont
  {P.}~\bibnamefont {Xue}},\ }\href {\doibase 10.1103/PhysRevA.107.L030201}
  {\bibfield  {journal} {\bibinfo  {journal} {Phys. Rev. A}\ }\textbf {\bibinfo
  {volume} {107}},\ \bibinfo {pages} {L030201} (\bibinfo {year}
  {2023})}\BibitemShut {NoStop}%
\bibitem [{\citenamefont {Chen}\ \emph {et~al.}(2010)\citenamefont {Chen},
  \citenamefont {Liu}, \citenamefont {Zhang},\ and\ \citenamefont
  {Wang}}]{PhysRevA.82.053841}%
  \BibitemOpen
  \bibfield  {author} {\bibinfo {author} {\bibfnamefont {Q.~H.}\ \bibnamefont
  {Chen}}, \bibinfo {author} {\bibfnamefont {T.}~\bibnamefont {Liu}}, \bibinfo
  {author} {\bibfnamefont {Y.~Y.}\ \bibnamefont {Zhang}}, \ and\ \bibinfo
  {author} {\bibfnamefont {K.~L.}\ \bibnamefont {Wang}},\ }\href {\doibase
  10.1103/PhysRevA.82.053841} {\bibfield  {journal} {\bibinfo  {journal} {Phys.
  Rev. A}\ }\textbf {\bibinfo {volume} {82}},\ \bibinfo {pages} {053841}
  (\bibinfo {year} {2010})}\BibitemShut {NoStop}%
\bibitem [{\citenamefont {Cirac}\ and\ \citenamefont
  {Zoller}(1995)}]{PhysRevLett.74.4091}%
  \BibitemOpen
  \bibfield  {author} {\bibinfo {author} {\bibfnamefont {J.~I.}\ \bibnamefont
  {Cirac}}\ and\ \bibinfo {author} {\bibfnamefont {P.}~\bibnamefont {Zoller}},\
  }\href {\doibase 10.1103/PhysRevLett.74.4091} {\bibfield  {journal} {\bibinfo
   {journal} {Phys. Rev. Lett.}\ }\textbf {\bibinfo {volume} {74}},\ \bibinfo
  {pages} {4091} (\bibinfo {year} {1995})}\BibitemShut {NoStop}%
\bibitem [{\citenamefont {Bruzewicz}\ \emph {et~al.}(2019)\citenamefont
  {Bruzewicz}, \citenamefont {Chiaverini}, \citenamefont {McConnell},\ and\
  \citenamefont {Sage}}]{Bruzewicz2019}%
  \BibitemOpen
  \bibfield  {author} {\bibinfo {author} {\bibfnamefont {C.~D.}\ \bibnamefont
  {Bruzewicz}}, \bibinfo {author} {\bibfnamefont {J.}~\bibnamefont
  {Chiaverini}}, \bibinfo {author} {\bibfnamefont {R.}~\bibnamefont
  {McConnell}}, \ and\ \bibinfo {author} {\bibfnamefont {J.~M.}\ \bibnamefont
  {Sage}},\ }\href {\doibase 10.1063/1.5088164} {\bibfield  {journal} {\bibinfo
   {journal} {Appl. Phys. Rev.}\ }\textbf {\bibinfo {volume} {6}},\ \bibinfo
  {pages} {021314} (\bibinfo {year} {2019})}\BibitemShut {NoStop}%
\bibitem [{\citenamefont {Georgescu}(2020)}]{Georgescu2020}%
  \BibitemOpen
  \bibfield  {author} {\bibinfo {author} {\bibfnamefont {I.}~\bibnamefont
  {Georgescu}},\ }\href {\doibase 10.1038/s42254-020-0189-1} {\bibfield
  {journal} {\bibinfo  {journal} {Nat. Rev. Phy.}\ }\textbf {\bibinfo {volume}
  {2}},\ \bibinfo {pages} {278} (\bibinfo {year} {2020})}\BibitemShut {NoStop}%
\bibitem [{\citenamefont {Lv}\ \emph {et~al.}(2018)\citenamefont {Lv},
  \citenamefont {An}, \citenamefont {Liu}, \citenamefont {Zhang}, \citenamefont
  {Pedernales}, \citenamefont {Lamata}, \citenamefont {Solano},\ and\
  \citenamefont {Kim}}]{PhysRevX.8.021027}%
  \BibitemOpen
  \bibfield  {author} {\bibinfo {author} {\bibfnamefont {D.}~\bibnamefont
  {Lv}}, \bibinfo {author} {\bibfnamefont {S.}~\bibnamefont {An}}, \bibinfo
  {author} {\bibfnamefont {Z.}~\bibnamefont {Liu}}, \bibinfo {author}
  {\bibfnamefont {J.-N.}\ \bibnamefont {Zhang}}, \bibinfo {author}
  {\bibfnamefont {J.~S.}\ \bibnamefont {Pedernales}}, \bibinfo {author}
  {\bibfnamefont {L.}~\bibnamefont {Lamata}}, \bibinfo {author} {\bibfnamefont
  {E.}~\bibnamefont {Solano}}, \ and\ \bibinfo {author} {\bibfnamefont
  {K.}~\bibnamefont {Kim}},\ }\href {\doibase 10.1103/PhysRevX.8.021027}
  {\bibfield  {journal} {\bibinfo  {journal} {Phys. Rev. X}\ }\textbf {\bibinfo
  {volume} {8}},\ \bibinfo {pages} {021027} (\bibinfo {year}
  {2018})}\BibitemShut {NoStop}%
\bibitem [{\citenamefont {Singha}\ \emph {et~al.}(2011)\citenamefont {Singha},
  \citenamefont {Gibertini}, \citenamefont {Karmakar}, \citenamefont {Yuan},
  \citenamefont {Polini}, \citenamefont {Vignale}, \citenamefont {Katsnelson},
  \citenamefont {Pinczuk}, \citenamefont {Pfeiffer}, \citenamefont {West},\
  and\ \citenamefont {Pellegrini}}]{A.2011}%
  \BibitemOpen
  \bibfield  {author} {\bibinfo {author} {\bibfnamefont {A.}~\bibnamefont
  {Singha}}, \bibinfo {author} {\bibfnamefont {M.}~\bibnamefont {Gibertini}},
  \bibinfo {author} {\bibfnamefont {B.}~\bibnamefont {Karmakar}}, \bibinfo
  {author} {\bibfnamefont {S.}~\bibnamefont {Yuan}}, \bibinfo {author}
  {\bibfnamefont {M.}~\bibnamefont {Polini}}, \bibinfo {author} {\bibfnamefont
  {G.}~\bibnamefont {Vignale}}, \bibinfo {author} {\bibfnamefont {M.~I.}\
  \bibnamefont {Katsnelson}}, \bibinfo {author} {\bibfnamefont
  {A.}~\bibnamefont {Pinczuk}}, \bibinfo {author} {\bibfnamefont {L.~N.}\
  \bibnamefont {Pfeiffer}}, \bibinfo {author} {\bibfnamefont {K.~W.}\
  \bibnamefont {West}}, \ and\ \bibinfo {author} {\bibfnamefont
  {V.}~\bibnamefont {Pellegrini}},\ }\href {\doibase 10.1126/science.1204333}
  {\bibfield  {journal} {\bibinfo  {journal} {Science}\ }\textbf {\bibinfo
  {volume} {332}},\ \bibinfo {pages} {1176} (\bibinfo {year}
  {2011})}\BibitemShut {NoStop}%
\end{thebibliography}%
\end{document}